\documentclass[12pt,preprint]{emulateapj}
\slugcomment{accepted by ApJ}

\shorttitle{Accretion in quiescent 
SMBHs: X-ray study}
\shortauthors{Soria et al.}

\begin{document}

%% LaTeX will automatically break titles if they run longer than
%% one line. However, you may use \\ to force a line break if
%% you desire.

\title{Accretion and nuclear activity of quiescent supermassive black
holes. I: X-ray study}

%% Use \author, \affil, and the \and command to format
%% author and affiliation information.
%% Note that \email has replaced the old \authoremail command
%% from AASTeX v4.0. You can use \email to mark an email address
%% anywhere in the paper, not just in the front matter.
%% As in the title, use \\ to force line breaks.

\author{R. Soria\altaffilmark{1}, G. Fabbiano\altaffilmark{1},
	Alister W. Graham\altaffilmark{2}, A. Baldi\altaffilmark{1}, 
	M. Elvis\altaffilmark{1},
	H. Jerjen\altaffilmark{2}, S. Pellegrini\altaffilmark{3},
	A. Siemiginowska\altaffilmark{1}}
\altaffiltext{1}{Harvard-Smithsonian Center for Astrophysics, 
	60 Garden st, Cambridge, MA 02138, USA}
\altaffiltext{2}{RSAA, Australian National University, Cotter Rd, ACT 2611}
\altaffiltext{3}{Astronomy Department, Bologna University, Italy}

%% Notice that each of these authors has alternate affiliations, which
%% are identified by the \altaffilmark after each name.  Specify alternate
%% affiliation information with \altaffiltext, with one command per each
%% affiliation.

%\altaffiltext{1}{Visiting Astronomer, Cerro Tololo Inter-American Observatory.
%`CTIO is operated by AURA, Inc.\ under contract to the National Science
%Foundation.}
%\altaffiltext{2}{Society of Fellows, Harvard University.}
%\altaffiltext{3}{present address: Center for Astrophysics,
%    60 Garden Street, Cambridge, MA 02138}
%\altaffiltext{4}{Visiting Programmer, Space Telescope Science Institute}
%\altaffiltext{5}{Patron, Alonso's Bar and Grill}

%% Mark off your abstract in the ``abstract'' environment. In the manuscript
%% style, abstract will output a Received/Accepted line after the
%% title and affiliation information. No date will appear since the author
%% does not have this information. The dates will be filled in by the
%% editorial office after submission.

\begin{abstract}
We have studied the nuclear activity in a sample 
of six quiescent early-type galaxies, 
with new {\it Chandra} data and archival {\it  HST} 
optical images. Their nuclear sources 
have X-ray luminosities $\sim 10^{38}$--$10^{39}$ erg s$^{-1}$ 
($L_{\rm X}/L_{\rm Edd} \sim 10^{-8}$--$10^{-7}$), 
and colors or spectra consistent with accreting supermassive 
black holes (SMBHs)---except for the nucleus of NGC\,4486B, which is 
softer than typical AGN spectra. 
In a few cases, the X-ray morphology of the nuclear sources 
shows hints of marginally extended structures, in addition 
to the surrounding diffuse thermal emission from hot gas, 
which is detectable on scales $\ga 1$ kpc.  
In one case (NGC\,5845), a dusty disk 
may partially obstruct our direct view of the SMBH.
We have estimated the temperature and density 
of the hot interstellar medium, 
which is one major source of fuel for the accreting SMBH; 
typical central densities 
are $n_e \approx (0.02 \pm 0.01)$ cm$^{-3}$.
Assuming that the hot gas is captured by 
the SMBH at the Bondi rate, we show that 
the observed X-ray luminosities are too faint to be consistent 
with standard disk accretion, but brighter than predicted 
by radiatively-inefficient solutions 
(e.g., advection-dominated acretion flows, ADAF). 
In total, there are $\approx 20$ galaxies 
for which SMBH mass, hot gas density, and 
nuclear X-ray luminosity are simultaneously known. 
In some cases, the nuclear sources are brighter than predicted 
by the ADAF model; in other cases, they are consistent 
or fainter. We discuss the apparent lack of 
correlations between Bondi rate and X-ray luminosity, 
and suggest that, in order to understand the observed 
distribution, we need to know two 
additional parameters: the amount of gas supplied by the  
stellar population inside the accretion radius, 
and the fraction (possibly $\ll 1$) 
of the total gas available that is accreted by the SMBH.
We leave a detailed study of these issues to a subsequent 
paper.
\end{abstract}

%% Keywords should appear after the \end{abstract} command. The uncommented
%% example has been keyed in ApJ style. See the instructions to authors
%% for the journal to which you are submitting your paper to determine
%% what keyword punctuation is appropriate.

%% Authors who wish to have the most important objects in their paper
%% linked in the electronic edition to a data center may do so in the
%% subject header.  Objects should be in the appropriate "individual"
%% headers (e.g. quasars: individual, stars: individual, etc.) with the
%% additional provision that the total number of headers, including each
%% individual object, not exceed six.  The \objectname{} macro, and its
%% alias \object{}, is used to mark each object.  The macro takes the object
%% name as its primary argument.  This name will appear in the paper
%% and serve as the link's anchor in the electronic edition if the name
%% is recognized by the data centers.  The macro also takes an optional
%% argument in parentheses in cases where the data center identification
%% differs from what is to be printed in the paper.


%% From the front matter, we move on to the body of the paper.
%% In the first two sections, notice the use of the natbib \citep
%% and \citet commands to identify citations.  The citations are
%% tied to the reference list via symbolic KEYs. The KEY corresponds
%% to the KEY in the \bibitem in the reference list below. We have
%% chosen the first three characters of the first author's name plus
%% the last two numeral of the year of publication as our KEY for
%% each reference.

\section{Introduction}

There is strong evidence that all massive galaxies 
(not including dwarfs) with a spheroidal component 
(ellipticals, lenticulars, and early-type 
spirals with bulges) host supermassive black holes (SMBHs) 
in their nuclei (Richstone et al.~1998). 
For a few dozen galaxies, there exist direct 
determinations of their nuclear masses, either from measurements 
of spatially-resolved stellar or gas kinematics (for quiescent 
nuclei) or from "reverberation mapping" of their broad-line regions 
(for active nuclei) (Gebhardt et al.~2003; Ferrarese et al.~2001; Ho 2002; 
Onken et al.~2004; Peterson et al.~2004). The masses of SMBHs correlate 
well with fundamental properties of their host galaxies: 
bulge mass (Magorrian et al.~1998), central velocity dispersion 
(Ferrarese \& Merritt 2000, Gebhardt et al. 2000), 
central light concentration (Graham et al.~2001) and 
mass of the dark-matter halo (Ferrarese~2002). 
However, this is not the case for the accreting luminosities 
of SMBHs: while a few are active (AGN), sometimes with 
bolometric luminosities close to or above their Eddington limits 
(Collin et al.~2002; Wang \& Netzer 2003), the majority are not. 
X-ray observations of these ``silent'' SMBHs 
have revealed faint sources with X-ray 
luminosities as low as $\sim 10^{-8} L_{\rm Edd}$
%\citep{fab97,ho99,ho01,pel05,fab03,fab04}.
(Fabbiano \& Juda 1997; Ho 1999; Ho et al.~2001; 
Pellegrini et al.~2003a; Pellegrini 2005; 
Fabbiano et al.~2003, 2004).

The bolometric luminosity can be expressed as $L_{\rm bol} 
\approx \eta \dot{M}
c^2$, where $\dot{M}$ is the mass accretion rate 
and $\eta$ is the efficiency of conversion of gravitational energy 
into radiation. It is still not clear whether the extremely 
low luminosity of most SMBHs is primarily due to 
a low accretion rate, or to a low radiative efficiency 
(i.e., $\eta \ll 0.1$); see Narayan (2002) 
for a review. If it is due to 
a low accretion rate, we need to determine whether 
this is caused by a lack of gas available in the inner 
few pc, or is instead the result of other phenomena 
(for example accretion disk winds or jets) that may prevent 
the gas from accreting onto the SMBH.

The most likely explanation appears to be that both low $\eta$ 
and low $\dot{M}$ play a part. 
It has become clear that a lack of fuel 
cannot be a general explanation: many elliptical 
galaxies do have large supplies of cold/warm interstellar gas 
at their centers (Caon, Macchetto \& Pastoriza 2000; 
Macchetto et al.~1996). 
%Even if this gas does not cool 
%efficiently, or does not have enough angular momentum to form 
%an accretion disk, 
Fabian \& Canizares (1988) pointed out that the accretion 
rate of the hot interstellar medium (ISM) expected 
from the standard Bondi (1952) model would still be high enough 
to produce X-ray luminosities at least two orders of magnitude 
higher than observed in the nuclei of 
most ellipticals, if accretion occurs efficiently ($\eta \sim 0.1$).
Recent high-resolution {\it Chandra} observations 
have led to the measurement of both 
the nuclear X-ray emission and of the 
physical parameters of the surrounding 
hot ISM, greatly strengthening  
this conclusion (e.g., Pellegrini 2005 and references 
therein).

Advection of energy into the SMBH 
(e.g., Narayan \& Yi 1994, 1995) is a plausible  
mechanism for low radiative efficiency, and may be 
the explanation for many extremely sub-Eddington sources.
However, it cannot be the only explanation 
for all sources, either. For example, advective accretion cannot 
be invoked to explain some very under-luminous stellar-mass 
compact objects in our Galaxy (Perna et al.~2003). Low $\dot{M}$ 
in addition to low $\eta$ has also been suggested as the major 
reason for the overall faintness of Sgr A* (Bower et al.~2003;
Quataert 2003). 

Hence, there is now considerable observational 
evidence that quiescent or very faint SMBHs 
are at least partly due to low radiative efficiency, 
and partly to accretion rates 
onto the compact object lower than predicted 
from the amount of gas available. However, 
the physical processes responsible 
for this reduced accretion rate are still not well understood, 
and various competing models have been proposed.
Hydrodynamical simulations (e.g., Perna et al.~2003, 
their Sect. 2.3 and references therein)
have shown that disk winds and outflows 
may prevent most of the inflowing gas 
from reaching the compact object, so that only a small 
fraction of the available mass is accreted. 
Another possibility is that convection sets up 
a barrier to accretion, by increasing the pressure 
and preventing the gas from falling in supersonically 
(Narayan, Igumenshchev, \& Abramowicz~2000). 
Alternatively, Tan \& Blackman (2005) speculated 
that efficient cooling 
inside the Bondi radius causes the infalling gas 
to condense into stars in a gravitationally unstable disk, 
instead of sinking into the SMBH.

Another promising scenario is that most 
of the accretion power is not radiated, 
but is instead carried away 
by a steady, relativistic jet, with bulk 
kinetic luminosity $L_{\rm J} \sim 0.1 \dot{M}c^2 
\gg \eta \dot{M}c^2$. In some cases 
(e.g., M\,87: di Matteo et al.~2003, IC\,4296: Pellegrini et al.~2003b; 
IC\,1459: Fabbiano et al.~2003), 
it has indeed been shown that including the kinetic 
energy of the jet reconciles the observed energy budget 
with the expected accretion rate.  If a relativistic 
jet is present, synchrotron radio emission is also expected. 
The ``fundamental-plane'' relation (Merloni, Heinz, \& Di Matteo 2003) 
between the observed radio luminosities, 
X-ray luminosities, and BH masses supports the hypothesis  
that inactive SMBHs have low accretion rates 
as well as low radiative efficiency ($\eta \la 10^{-3}$), 
and that most of the accretion power is carried outwards 
by a steady jet.
However, in other cases (e.g., NGC\,821: Fabbiano et al.~2004) 
it is more likely that the accretion power is released 
in intermittent, short-lived outbursts (Binney \& Tabor 1995; 
Ciotti \& Ostriker 2001; Jones et al.~2002; Omma et al.~2004); 
if that is the case, the relative fraction of SMBHs observed 
in an active or quiescent state 
would depend on the duration of the duty cycle.
%(Binney & Tabor 1995, MNRAS, 276, 663; Ciotti & Ostriker 2001, 2001, ApJ, 551, 131; Jones et al. 2002, ApJ 567, L115; Omma et al. 2004, MNRAS 348, 1105).
Alternatively, in a few cases the X-ray emission from the SMBH 
may be absorbed by dust (often found in the inner $\sim 100$ pc 
of elliptical galaxies: Rest et al.~2001) and re-emitted in the infrared (IR).
%A dusty disk is often found in the inner $\sim 100$ pc 
%of elliptical galaxies, especially those 
%with disky isophotes (van den Bosch et al.~1994). 
Finally, it is still possible that a few luminous 
elliptical galaxies with partially depleted stellar cores 
may not have large gas supplies to feed the SMBH.
Determining which of these physical processes 
are at play in quiescent SMBHs will also 
help us understand and quantify the SMBH feedback 
onto its host galaxy.
%; in particular, 
%whether the accretion energy is carried out by a jet, 
%or whether there are intermittent outbursts.

For a sample of 18 nearby galaxies with nuclear 
X-ray luminosity $L_{\rm X}$ and Bondi mass accretion rate 
of hot gas estimated from {\it {Chandra}} data, 
Pellegrini (2005) found that $L_{\rm X}$ can vary
by orders of magnitude and shows 
a large scatter when plotted against
BH mass or Bondi rate. While in some cases 
the observed luminosities are consistent with the prediction 
of advective models, for other nuclear sources 
the observed luminosities are much lower, and 
therefore scenarios including outflows or convection 
are needed. A scenario with feedback from accretion
on the surrounding gas was considered the most promising.

\subsection{Objectives and targets of our study}

To test these scenarios, we chose a further 
sample of {\it six nearby, 
quiescent elliptical galaxies} with direct SMBH mass determinations, 
from Gebhardt et al.~(2003) and Kormendy et al.~(1997). 
Stringent upper limits on AGN activity exist for all these SMBHs, 
from radio, optical (Balmer emission lines), 
and previous X-ray studies (Ho 2002), and from 
a lack of UV bumps. Our goal is to explore the most extreme cases 
of sub-Eddington nuclear activity, so we only focused on galaxies 
with SMBH masses $> 10^7 M_{\odot}$ and X-ray luminosity $< 10^{40}$ 
erg s$^{-1}$. We also excluded galaxies more distant than 30 Mpc. 

In five of the six galaxies, the SMBH mass has been accurately 
determined with three-integral models 
(Gebhardt et al.~2003)\footnote{In three of those five nuclei, 
the dynamical sphere of influence of the SMBH 
has been resolved by {\it HST} 
(Merritt \& Ferrarese 2001; Ho 2002), 
making the mass determination more reliable; in the remaining 
two cases, NGC\,821 and NGC\,3377, the sphere of influence 
is not resolved.}.
The only exception is NGC\,4486B, whose mass estimate 
was based on two-integral models (Kormendy et al.~1997) 
and is therefore much more uncertain. For example, 
a revised error range in Kormendy \& Gebhardt (2001) 
allows for masses between $2 \times 10^7$ and $10^9 M_{\odot}$. 
A mass $\approx 7 \times 10^8 M_{\odot}$ was 
quoted as a preliminary result of an ongoing  
study by Gebhardt (2005, priv.~comm.).
As a corollary of our study, we have obtained 
an alternative (indirect) SMBH mass 
determination for NGC\,4486B 
(Soria et al.~2005 = Paper II), using 
the $M_{\rm BH}$--velocity dispersion 
(Tremaine et al.~2002)
and $M_{\rm BH}$--S\'{e}rsic index 
(Graham et al.~2003, 2005 in prep.) relations.
%It has since been suggested that the published 
%value overestimates the real mass 
%by an order of magnitude (D. Merritt, priv.~comm.).
We are aware that the larger mass uncertainty, 
and its peculiar morphological type, make NGC\,4486B 
somewhat different from the rest of our {\it Chandra} targets, 
among which it was included for historical reasons.
Classified as a compact elliptical (cE) galaxy, 
NGC\,4486B may have originated 
as the bulge of a much larger, 
tidally stripped disk galaxy 
(Jerjen \& Binggeli 1997; see also Graham 2002).

Our main objective is to {\em determine the energy 
and mass budget} for the nuclear regions of 
our sample galaxies, and  
for a few others for which SMBH mass and luminosity 
are available in the literature. 
We want to answer three 
fundamental questions: how much gas is available for accretion? 
how much energy is radiated by the SMBH?
where does the rest of the mass and energy go? 
Moreover, this may also help 
to explain why most of the 
SMBHs in the local universe are very faint or quiescent.

We begin by using {\it Chandra} data to investigate 
the spatial and spectral properties of the nuclear 
X-ray emission. By comparing the observed X-ray luminosity 
of the SMBH with the temperature and density 
of the surrounding gas, we can constrain some of 
the accretion parameters. In particular, if we 
could precisely determine the mass accretion rate 
onto the SMBH, we would find the efficiency.
However, a direct measurement of the accretion rate 
is still beyond today's observational capabilities, 
given the characteristic size of SMBH event 
horizons ($\sim 10^{-5}$ pc). 
So, we need to turn the problem around. 
We shall use model predictions for the radiative 
and non-radiative efficiency (e.g., standard disks, 
or advective accretion) and compare them with 
the observed luminosity and gas density available 
inside the SMBH accretion radii (characteristic 
size $\sim 10$ pc). From this, we will infer 
what fraction of the gas available is really 
accreted. In Paper I, we shall use {\it Chandra} 
to estimate the hot gas component to the accretion 
flow; we shall look for correlations between 
the classical Bondi inflow rate from the hot ISM, 
and the observed SMBH luminosity.
In Paper II, we shall use optical 
data to determine whether additional sources 
of fuel (e.g., warm gas from stellar winds) 
may have been missed by the X-ray study.
We shall then discuss the correlation between 
the total gas accretion rate and the observed 
SMBH luminosities.
Finally, we shall discuss how to balance the 
mass and energy budget, taking into account 
total gas injection, accretion onto the SMBH, 
and outflows.

%%%\clearpage
%\begin{table}
%\begin{center}
%\caption{Targets of our study \label{tbl-1}}
%\begin{tabular}{lcccccr}
\begin{deluxetable*}{lccccccc}
\tabletypesize{\scriptsize}
%\rotate
\tablecaption{Targets of our study \label{tbl-1}}
\tablewidth{0pt}
\tablehead
{
%\colhead{\,}&\colhead{\ }&\colhead{}&\colhead{}
%	&\colhead{\ }&\colhead{}&\colhead{}&\colhead{}\\[8pt]
\colhead{Galaxy} & \colhead{Type} & \colhead{d} & \colhead{$M_B$} 
	& \colhead{$M_{\rm BH}$}
	& \colhead{Obs. Date} & \colhead{Obs. ID}  
	& \colhead{Exp.~time} \\[2pt]
   	& & \colhead{(Mpc)} & \colhead{(mag)} & 
	\colhead{($10^8 M_{\odot}$)} & & & \colhead{(ks)}\\[2pt]
\colhead{(1)}  &  \colhead{(2)}  &  \colhead{(3)}  & \colhead{(4)}  &  
                  \colhead{(5)}  &  \colhead{(6)}  &  \colhead{(7)} 
		 &  \colhead{(8)}}
\startdata
\, & \, & & & & & & \\[6pt]
NGC\,821 
	& E6 & $24.1 \pm 2.0$ & $-20.7$ & $0.85^{+0.35}_{-0.35}$ 
	& 2002/11/26 & 4408 & 25.3\\
 & & & & & 2002/12/01 & 4006 & 13.7\\[4pt]
NGC\,3377 & E5-6& $11.2 \pm 0.5$ & $-19.2$ & $1.0^{+0.9}_{-0.1}$ 
	& 2003/01/06 & 2934 & 39.6\\[4pt]
NGC\,4486B & cE0 & $16.9 \pm 1.3$ & $-16.8$ & 
	$\left[6.0^{+3.0}_{-2.0}\right]$ 
	& 2003/11/21 & 4007 & 36.2\\[4pt]
 &&&& $0.5^{+0.5}_{-0.2}$ &&\\[4pt]
NGC\,4564 & E6/S0 & $15.0 \pm 1.2$ & $-19.0$ & $0.56^{+0.03}_{-0.08}$ 
	& 2003/11/21 & 4008 & 17.9\\[4pt]
NGC\,4697 
	& E6 & $11.7 \pm 0.8$ & $-20.3$&$1.7^{+0.2}_{-0.1}$ 
	& 2000/01/15 & 784 & 39.3\\
 & & & & & 2003/12/26 & 4727 & 39.9\\
 & & & & & 2004/01/06 &  4728 & 35.7\\
 & & & & & 2004/02/12 &  4729 & 38.1\\[4pt]
NGC\,5845 & E3 & $25.9 \pm 2.7$ & $-18.8$& $2.4^{+0.4}_{-1.4}$
	& 2003/01/03 & 4009 & 30.0\\[4pt]
\enddata
\tablecomments{
Col.(1): see Fabbiano et al.~(2004) for a detailed 
 study of NGC\,821, and Sarazin, Irwin \& Bregman (2001) for the analysis
of the first observation of NGC\,4697.
Col.(2): from the NASA/IPAC Extragalactic Database (NED).
Col.(3): from Tonry et al.~(2001), except for NGC\,4486B,
	from Neilsen \& Tsvetanov (2000).
Col.(4): from NED, adopting the distances in Col.(3).
Col.(5): from Gebhardt et al.~(2003), except for NGC\,821 
        (Richstone et al.~2004), and for NGC\,4486B. 
	The latter has two alternative values. The higher mass 
	(in square brackets)
	is from Kormendy et al.~(1997), using a two-integral model.
	A similar result ($5.0^{+4.9}_{-4.8} \times 10^8 M_{\odot}$) 
	was obtained with the same method by Kormendy \& Gebhardt
	(2001). However, the reliability of this value 
	is questionable, both because of the method on which 
	it is based, and because it is uniquely discrepant 
	from the BH mass inferred from all indirect indicators 
	(i.e., the global correlations with stellar velocity 
	dispersion, S\'ersic index, bulge mass).	
	The alternative, indirect mass determination  
	was obtained in our Paper II (Soria et al.~2005), from 
	the global correlation between S\'ersic index 
	in the optical brightness profile 
	and BH mass (Graham et al.\ 2003, 2005 in prep.).
}
%Take a look in Kormendy & Gebhardt (2001, astro-ph/0105230)
%Their Table 1 gives the mass as 0.2 - 9.9 x 108 M_sun
%A huge range, with the optimal value at 5.0 x 108
%\tablenotetext{b}{Yet another sample footnote for table~\ref{tbl-2}}
%\tablenotetext{c}{Another sample footnote for table~\ref{tbl-2}}
%\tablecomments{We can also attach a long-ish paragraph of explanatory
%material to a table.}
%\end{center}
\end{deluxetable*}
%%%%\clearpage

\section{X-ray Observations and Data analysis}

The six elliptical galaxies that are the 
object of our study (Table 1) were 
observed with {\it Chandra} (Weisskopf et al.~2000) 
ACIS-S, centered on chip 7 (i.e., ACIS-S3). 
For one of the galaxies, NGC\,4697, we used all 4 observations 
currently available in the public archive.
We analysed the data with the {\footnotesize {CIAO}} v3.2 
software\footnote{http://cxc.harvard.edu/ciao/} 
for NGC\,4697, and with {\footnotesize {CIAO}} v3.1 for the 
other targets; we checked that the difference between 
the two {\footnotesize {CIAO}} versions is negligible 
for the purposes of this work.
%We then used {\footnotesize {XSPEC}} 11.3.1 for the spectral fitting.
We downloaded the bias files for each observations, 
and built new bad-pixel files\footnote{Thread at 
http://cxc.harvard.edu/ciao/threads/acishotpixels/}. 
We then 
built new level=1 and level=2 event files with the {\footnotesize {CIAO}} 
script {\tt {acis\_process\_events}}, which includes the application 
of a time-dependent gain correction to the data. (The time-dependent 
gain adjustment was not applied to the 2000 observation 
of NGC\,4697, because it was taken with a focal plane temperature 
different from the normal $-120$ C).
We also improved the astrometric accuracy by calculating 
and removing the (negligible) aspect offsets, 
using the {\footnotesize {CIAO}} Aspect 
Calculator\footnote{Thread at 
http://cxc.harvard.edu/ciao/threads/arcsec\_correction/}.
No significant background flares were observed in these data, 
so no time filtering was necessary. The background level 
was generally very low except for the observation of NGC\,4486B.
The energy range was filtered to $0.3$--$10$ keV, with 
the {\footnotesize {CIAO}} task {\tt {dmcopy}}.
We identified the point sources with the {\tt {wavdetect}} task, 
and for the brightest of them, we extracted their spectra 
with {\tt {psextract}} (which also builds response 
and auxiliary response files, 
with the {\tt {mkrmf/mkarf}} tasks).
After subtracting the discrete sources, we used the 
{\footnotesize{CIAO}} script {\tt{acisspec}} to extract 
the spectra of diffuse emission from extended regions, 
together with their weighted response files and backgrounds.
We then used {\footnotesize {XSPEC}} 11.3.1 (Arnaud 1996) 
for the spectral fitting of point-like and extended 
sources\footnote{http://heasarc.gsfc.nasa.gov/lheasoft/xanadu/xspec/}.

\section{Results}

We focus in particular on the new results 
for NGC\,3377, NGC\,4486B, NGC\,4564 and NGC\,5845.
The remaining two galaxies have already been studied 
in detail (NGC\,821: Fabbiano et al.~2004; 
NGC\,4697: Sarazin et al.~2001): here we re-analyse 
and discuss their nuclear X-ray properties 
in the context of our new sample of massive, 
quiescent elliptical galaxies. In particular, 
for NGC\,4697 we use both the 2000 {\it Chandra} 
observation that was studied by Sarazin et al.~(2001), 
and three more recent observations that became 
public in early 2005.

%%%\clearpage
\begin{deluxetable*}{l@{\ \ \ \ }r@{\ \ }rrr@{\ \ \ \ }r@{\ \ }rrr}
\tabletypesize{\scriptsize}
%\rotate
\tablecaption{ACIS-S net counts and count rates from the nuclear sources\label{tbl-2}}
\tablewidth{0pt}
\tablehead
{
%\begin{table*}
%\begin{center}
%\caption{ACIS-S net counts and count rates from the nuclear sources\label{tbl-2}}
%\tableline\tableline
\colhead{Galaxy} & \multicolumn{4}{c}{ACIS-S Net Counts} 
	& \multicolumn{4}{c}{ACIS-S Count Rates ($10^{-4}$ s$^{-1}$)}\\
 & $0.3$--$10$ keV & $0.3$--$1$ keV & $1$--$2$ keV & $2$--$10$ keV & 
	$0.3$--$10$ keV & $0.3$--$1$ keV & $1$--$2$ keV & $2$--$10$ keV  }
%\tableline\\[-3pt]
\startdata
\, & \, & & & & & & &\\[-3pt]
NGC\,821 (core) &\multicolumn{8}{c}{not detected}\\
NGC\,821 (``jet'')
	& $137.2\pm26.4$ & $34.3 \pm 7.7$ & $63.5\pm9.5$ & $39.4\pm23.4$
	 &$36.2 \pm 7.0$ &$9.0 \pm 2.0$ & $16.8 \pm 2.5$ &$10.4 \pm 6.2$\\[2pt]
NGC\,3377 & $113.1\pm11.6$ & $41.5 \pm 6.8$ & $45.5 \pm 6.8$ & $26.1\pm 5.4$&
	$29.3 \pm 3.1$ & $10.9 \pm 1.8$ & $11.5 \pm 2.4$ & $6.9 \pm 1.4$\\[2pt]
NGC\,4486B & $38.2\pm8.2$ & $27.7 \pm 6.2$ & $6.8 \pm 7.0$ &$3.7\pm 5.1$
	& $10.6 \pm 2.3$ &$7.7 \pm 1.7$ & $1.9 \pm 1.9$ & $1.0 \pm 1.4$\\[2pt]
NGC\,4564 &$72.3\pm9.3$ & $28.3 \pm 5.9$ & $27.5 \pm 5.6$ &$16.5\pm 4.4$ 
	& $43.7 \pm 5.6$ &$17.0 \pm 3.5$ & $16.7 \pm 3.4$ & 
	$10.0 \pm 2.7$\\[2pt]
NGC\,4697 (1)
	&$128.9\pm11.6$ & $57.3 \pm 7.8$ & $40.4 \pm 6.4$ &$31.4\pm 5.7$ 
	& $32.8 \pm 3.0$ &$14.6 \pm 2.0$ & $10.3 \pm 1.6$ & $8.0 \pm
1.4$\\ 
NGC\,4697 (2)
	&$288.0\pm17.3$ & $97.0 \pm 10.2$ & $120.7 \pm 11.1$ &$69.3\pm 8.5$ 
	& $25.3 \pm 1.5$ &$8.5 \pm 0.9$ & $10.6 \pm 1.0$ & $6.1 \pm 0.7$\\[2pt]
NGC\,5845 &$117.8\pm11.8$ & $36.5 \pm 6.8$ & $56.9 \pm 7.9$ &$24.4\pm 5.3$ 
	& $39.4 \pm 3.9$ &$12.2 \pm 2.3$ & $19.0 \pm 2.7$ & $8.2 \pm 1.8$ \\
\enddata
\tablecomments{
%\tableline
%\end{tabular}
%\tablenotetext{a}{}
The values for NGC\,821 are from the combined 2002 Nov/Dec
observations; see Fabbiano et
al.~(2004) for details on the definition of the jet-like 
nuclear feature as opposed to a point-like (non-detected) core. The
first line for NGC\,4697 is from the 2000 Jan observation, and the second 
is for the combined 2003 Dec--2004 Feb observations; the reduction 
in count rate is due to the degraded performance of ACIS-S.} 
%\tablenotetext{b}{}
%\tablenotetext{c}{}
%\tablecomments{We can also attach a long-ish paragraph of explanatory
%material to a table.}
%\end{tabular}
%\end{center}
%\end{table*}
\end{deluxetable*}
%%%\clearpage

\subsection{Morphology of the nuclear X-ray sources}

In all six galaxies we detect X-ray emission at and around 
the nuclear position. In five cases (all except NGC\,821, 
see Fabbiano et al.~2004), the centroid 
of the nuclear X-ray source coincides with 
the optical/IR nucleus identified in the 
Two-Micron All Sky Survey (2MASS) 
and United States Naval Observatory (USNO) 
Catalogues, within the astrometric 
accuracy of {\it Chandra} (the $90\%$ uncertainty circle
has a radius of $\la 0\farcs6$). 
For NGC\,4697, relative astrometry with an accuracy 
better than $\approx 0\farcs3$ can be obtained, 
because a few non-nuclear X-ray sources have 
point-like optical counterparts (globular clusters).

Firstly, we analyzed the morphology of the nuclear X-ray sources 
to determine whether they are consistent 
with point-like emission from the SMBH 
or are instead extended. From the theoretical ACIS-S3 
point spread function and from our fitting of other point 
sources detected in the central region of the chip, 
we expect a Gaussian FWHM $= 0\farcs89 \pm 0\farcs15$ 
for a point-like source in the $0.3$--$8$ keV energy band.
%We find that, for point-like sources in the ACIS-S3 chip, 
%the gaussian FWHM $= 0\farcs89 \pm 0\farcs15$.
The (weak) nuclear source of NGC\,4486B is consistent 
with being point-like (FWHM $\approx 0\farcs9$), 
implying a size $\la 70$ pc; 
it is also well isolated: no other X-ray 
sources are detected in that galaxy.
The nuclear source in NGC\,4697 (Figure~1, bottom left) 
is also point-like (size $\la 50$ pc) but is located 
in a region with high source density, 
with at least another 5 point-like sources 
detected within a radius of $300$ pc. Their 
spatial distribution around the nuclear source is probably 
a chance grouping of X-ray binaries in the galactic bulge.

%may be extended, with a FWHM $= 1\farcs15 \approx 65$ pc;
The nuclear source of NGC\,4564 (Figure~1, top right) 
has a FWHM $\approx 2\farcs2 \approx 160$ pc 
in the $0.3$--$8$ keV band; therefore, 
it is probably extended.
In the remaining three galaxies, NGC\,3377, NGC\,5845 
and NGC\,821, structure in the nuclear region 
is more clearly visible. In NGC\,5845 (Figure~1, 
bottom right), the nuclear emission has 
an irregular, ``fuzzy'' appearance over 
a region $\approx 600$ pc across. In the other 
two galaxies, NGC\,821 and NGC\,3377, the emission 
has instead an elongated, ``jet-like'' appearance.
In NGC\,3377 (Figure~1, top left), an elongated feature 
seems to link the nuclear source 
with another, fainter source located $\approx 3\arcsec 
\approx 160$ pc to the East. For a discussion 
of the elongated substructure in the 
nucleus of NGC\,821, see Fabbiano et al.~(2004).
Limiting our imaging study to the $1.5$--$8$ keV band 
(Figure 2), we still find some extended structure 
in NGC\,5845.

\subsection{X-ray colors and spectra of the nuclear sources}

We extracted X-ray spectra of the nuclear sources, 
using the extraction regions plotted in Figure~1 
(dashed white circles); the background regions 
were chosen as annuli around the nuclear sources, 
and we took care of excluding nearby sources.
For NGC\,4697, we extracted separate spectra 
for the four observations. For some tasks, it is possible 
and convenient to coadd the 2003 Dec--2004 Feb 
observations and study them together; however, 
the analysis of the first dataset (2000 Jan) 
has to be carried out separately because 
of the ACIS-S sensitivity degradation, particularly 
at lower energies, over those four years.

Firstly, we compared the broad-band X-ray colors (Table 2), 
before introducing any spectral modelling.
In five of the six nuclear sources, the X-ray colors 
are consistent with typical AGN spectra, 
a power-law spectrum of index $\approx 1.5$--$2$ (Figure 3).
The only exception is NGC\,4486B, which has softer 
colors, with most of the counts detected below 1 keV, 
although the errors are large because 
of the relatively high background level and the small 
number of counts. 

We then estimated the emitted fluxes and luminosities 
by assuming an absorption column density equal 
to the Galactic line-of-sight (Table 3; from 
Dickey \& Lockman 1990). The color-color 
diagram (Figure 3) and the X-ray spectra (at least for the nuclear 
sources with enough counts) justify this assumption, 
suggesting that the total column density 
is not much higher than the Galactic value; 
hence, we are only slightly underestimating the emitted 
luminosities. We fitted the (ungrouped) spectra 
with an absorbed power-law (Table 3), using the Cash statistic 
(Cash 1979)\footnote{This statistic 
can be used in the same way as the $\chi^2$ statistic 
to find confidence intervals. One finds the parameter 
values that give C $=$ C$_{\rm min}$ $+$ N, where N 
is the same number that gives the required 
confidence for the number of interesting parameters 
as for the $\chi^2$ case (Arnaud 1996). 
See the XSPEC manual for details.}. 
We also grouped the same spectra to achieve 
a minimum signal-to-noise ratio of $3$ in each spectral 
bin, and fitted them with the $\chi^2$ statistics 
(Table 3). For sources with a very limited number of counts, 
we consider the results based on the 
Cash statistic to be more reliable.
For NGC\,4697, we increased the signal-to-noise ratio 
by coadding the spectra from 2000 and 
2003--2004, using the algorithm of 
Page, Davis \& Salvi (2003), which generates 
a count-weighted average of their response matrices 
(Figure~4). From a power-law fit (Table 3) 
we calculate an emitted luminosity 
$L_{\rm x} = (3.9\pm0.2) \times 10^{38}$ erg s$^{-1}$ 
in the $0.3$--$10$ keV band; we also checked that 
there are no significant spectral or luminosity changes 
between the two epochs.
The emitted luminosities of the nuclear sources 
in the other five galaxies are between $\approx 3 \times 10^{38}$ 
and $\approx 3 \times 10^{39}$ erg s$^{-1}$, 
in the same band (Table 3).

For NGC\,4486B, we have also tried fitting 
the nuclear source with a thermal-plasma 
model, to account for the possibility 
that it is a very compact patch of thermal 
emission from the ISM. 
We obtain a best-fit $kT = 3.7^{+8.0}_{-1.6}$ 
keV, but the fit is statistically worse 
(C-statistics $=54/297$) than 
for the power-law model (C-statistics $=46/297$). 
Henceforth, we shall assume for this work 
that most of the X-ray emission comes from the SMBH.

%%%\clearpage
\begin{figure*}[t]
\epsscale{1.0}
\begin{center}
\includegraphics[angle=0,scale=0.37]{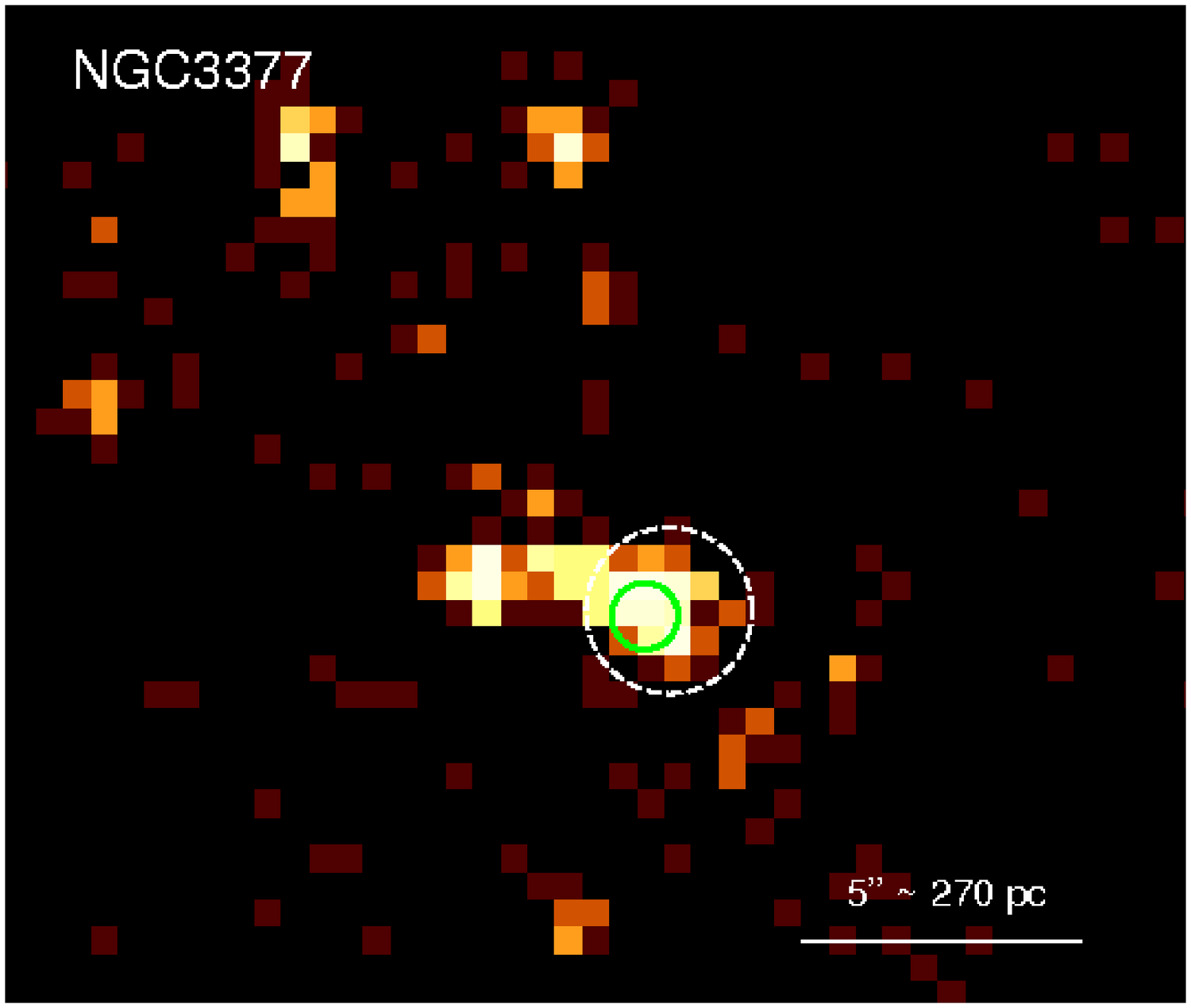}
\includegraphics[angle=0,scale=0.37]{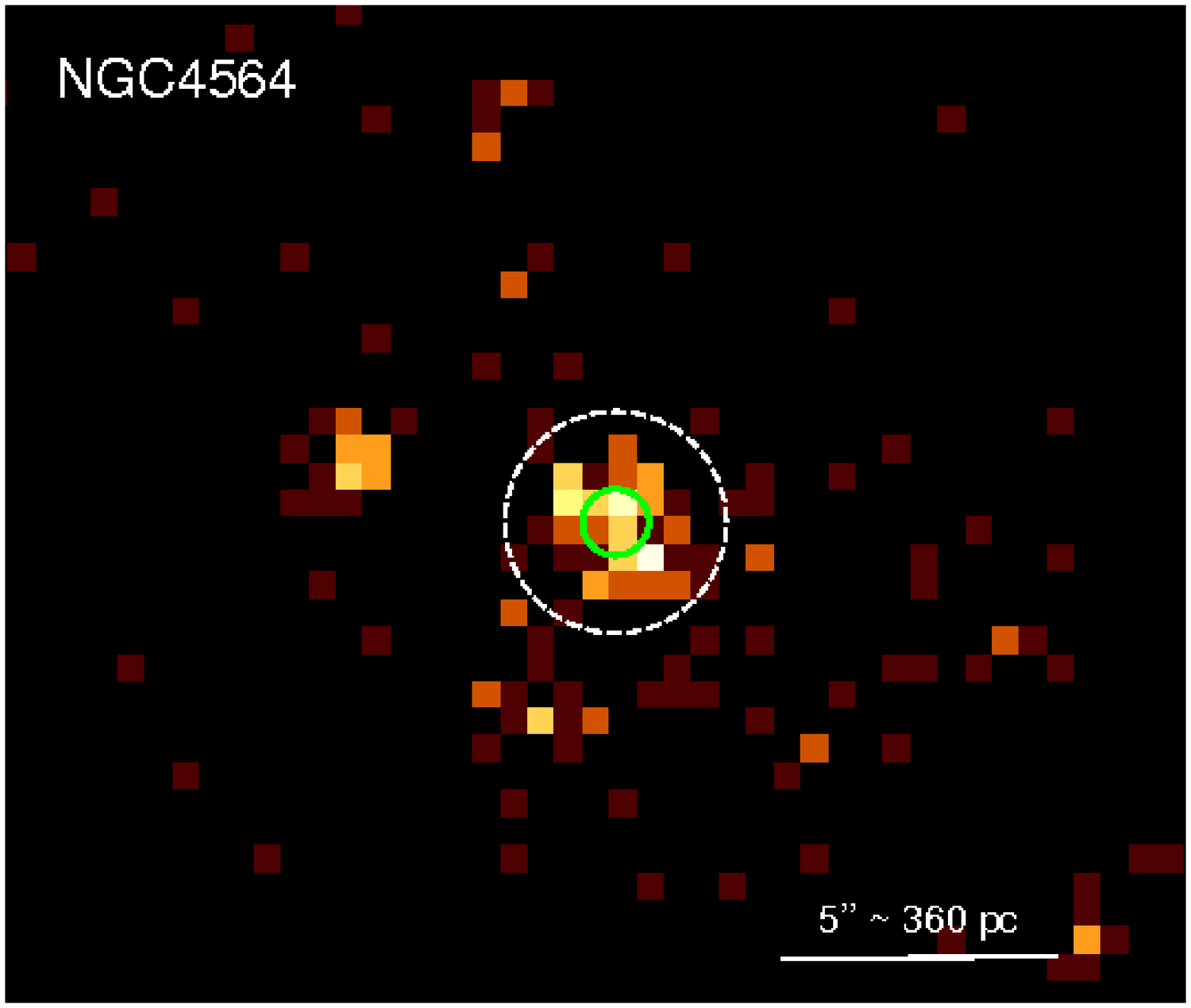}\\
\includegraphics[angle=0,scale=0.37]{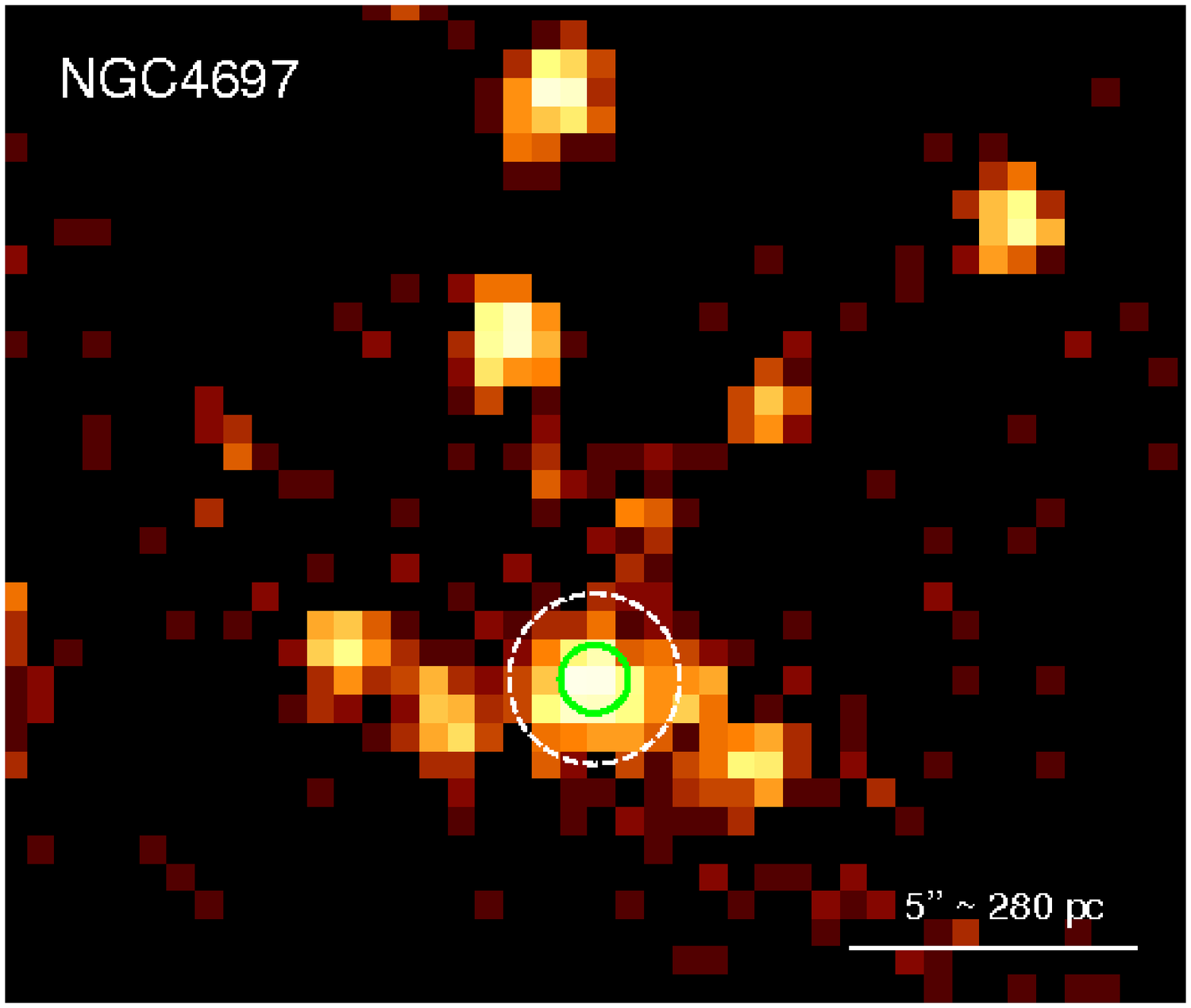}
\includegraphics[angle=0,scale=0.365]{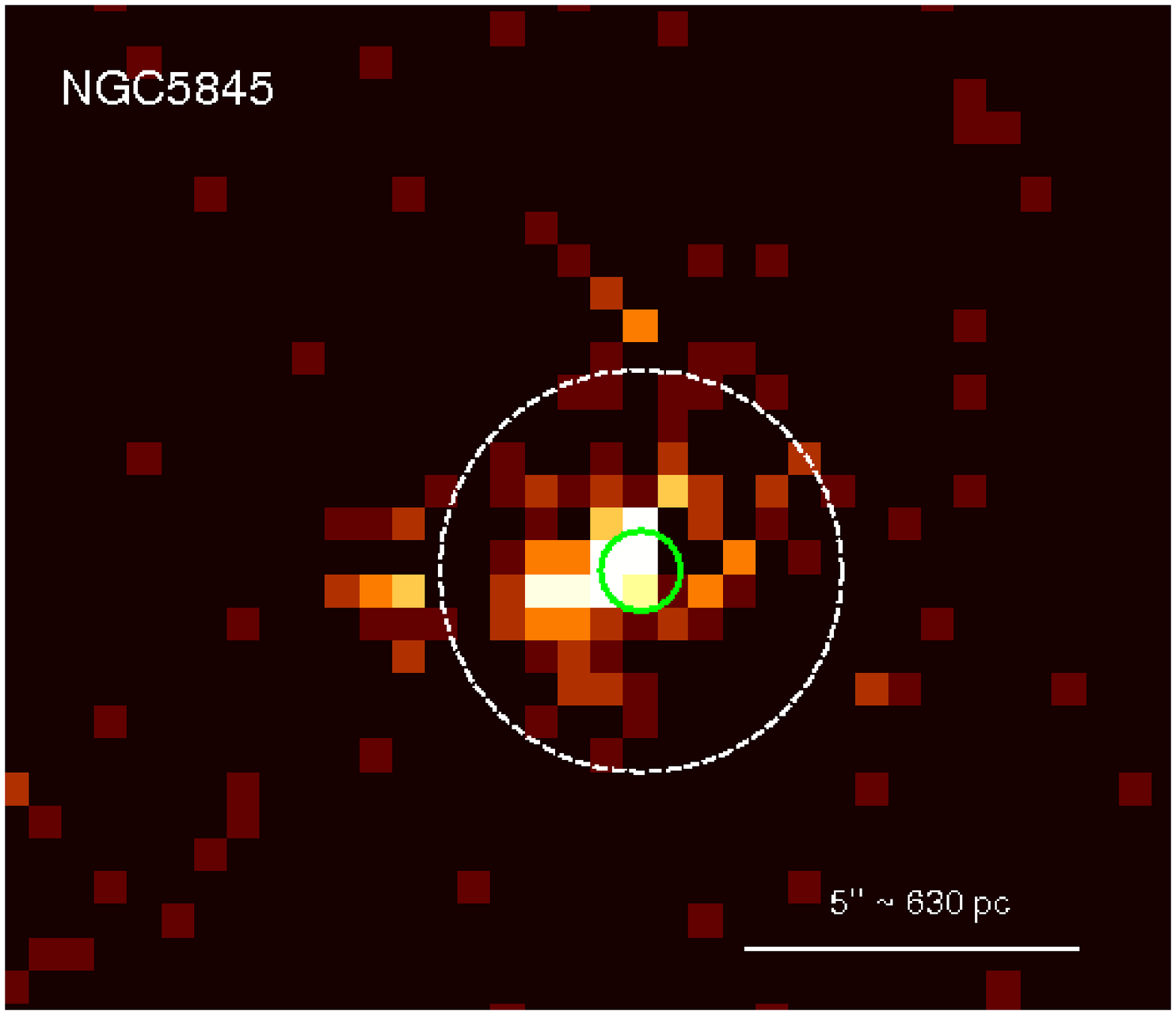}
\end{center}
%\plotone{acis4.ps}
\caption{ACIS-S image ($0.3$--$8$ keV band) 
of the nuclear regions in four 
of the galaxies showing extended structures.  (For the other 
two galaxies in our sample: see Fabbiano et al.~(2004) for 
NGC\,821; the nuclear source is point-like in NGC\,4486B).
In each panel, the dashed white circle is the source region 
used to determine the nuclear luminosity, spectrum 
and colors (Tables 2, 3 and Figure 3). The green circles 
(radius of $0\farcs6$) are the 2MASS positions 
of the galactic nuclei, except for NGC\,4697, where, 
for consistency with previous work, the green circle 
marks the average between the 2MASS and the USNO B1.0 
Catalogue positions (the difference between 
the two is $0\farcs4$). North is up, East to the left.
\label{fig1}}
\end{figure*}

\begin{figure*}[t]
\epsscale{1.0}
\begin{center}
\includegraphics[angle=0,scale=0.37]{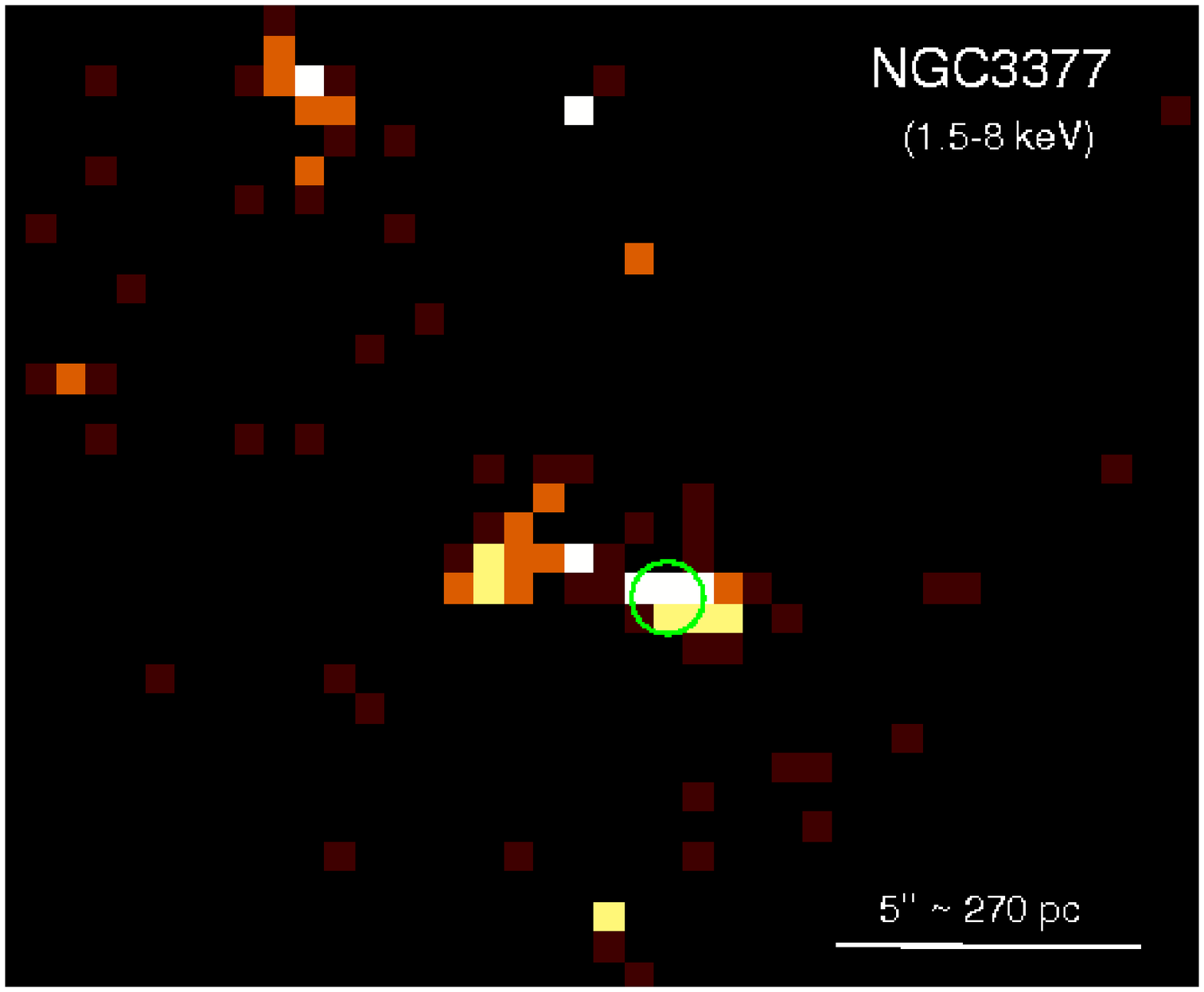}
\includegraphics[angle=0,scale=0.37]{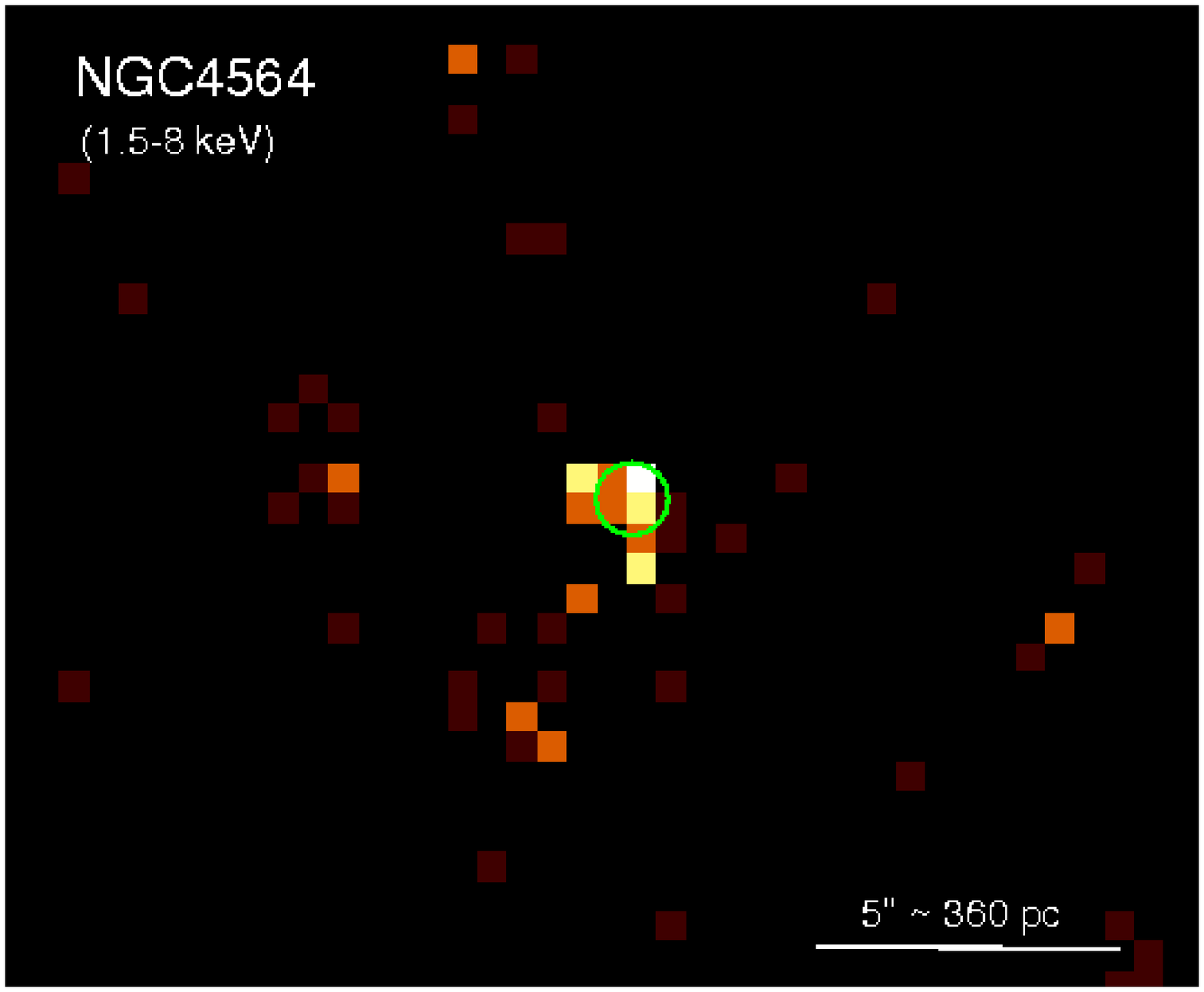}\\
\includegraphics[angle=0,scale=0.37]{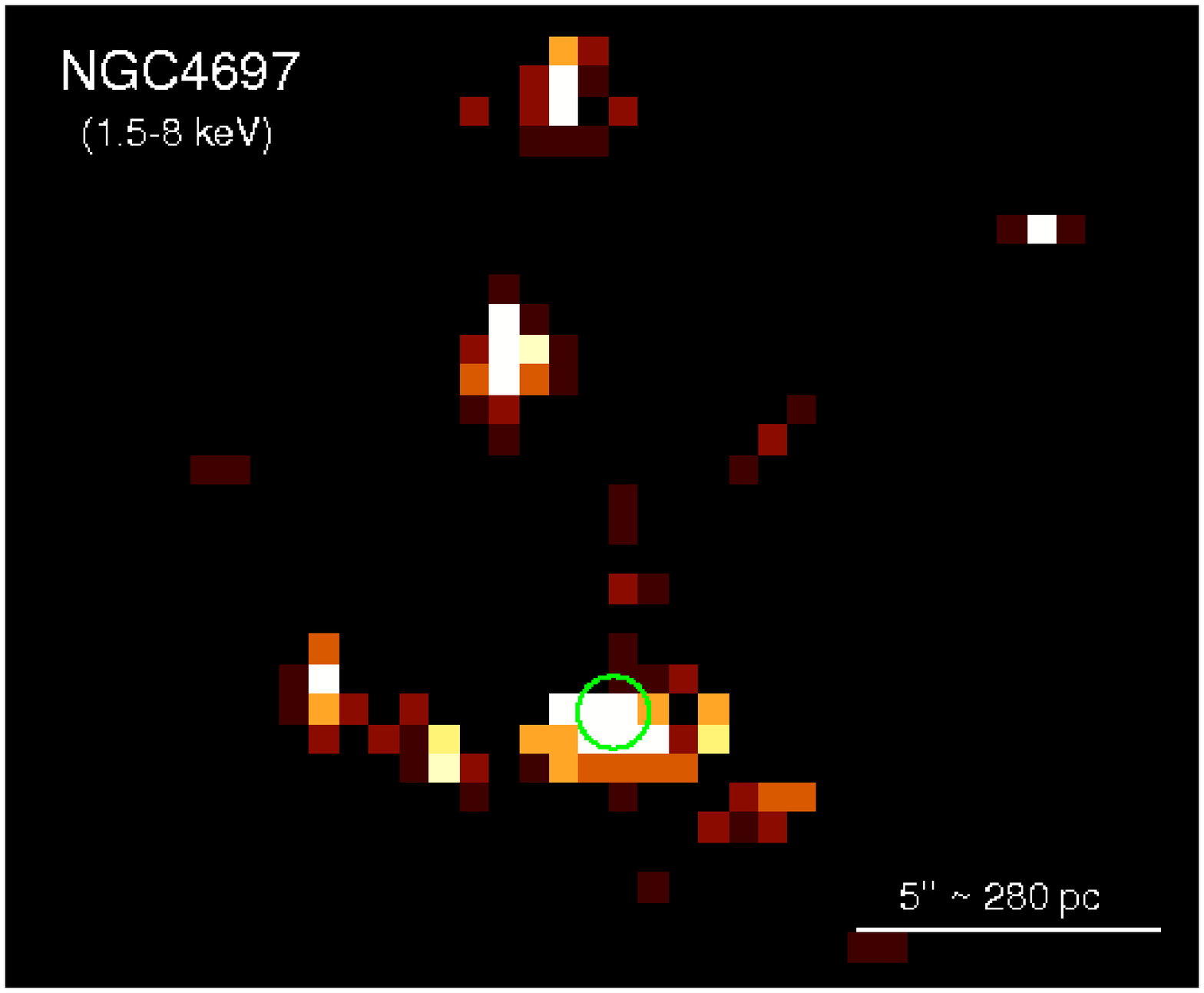}
\includegraphics[angle=0,scale=0.37]{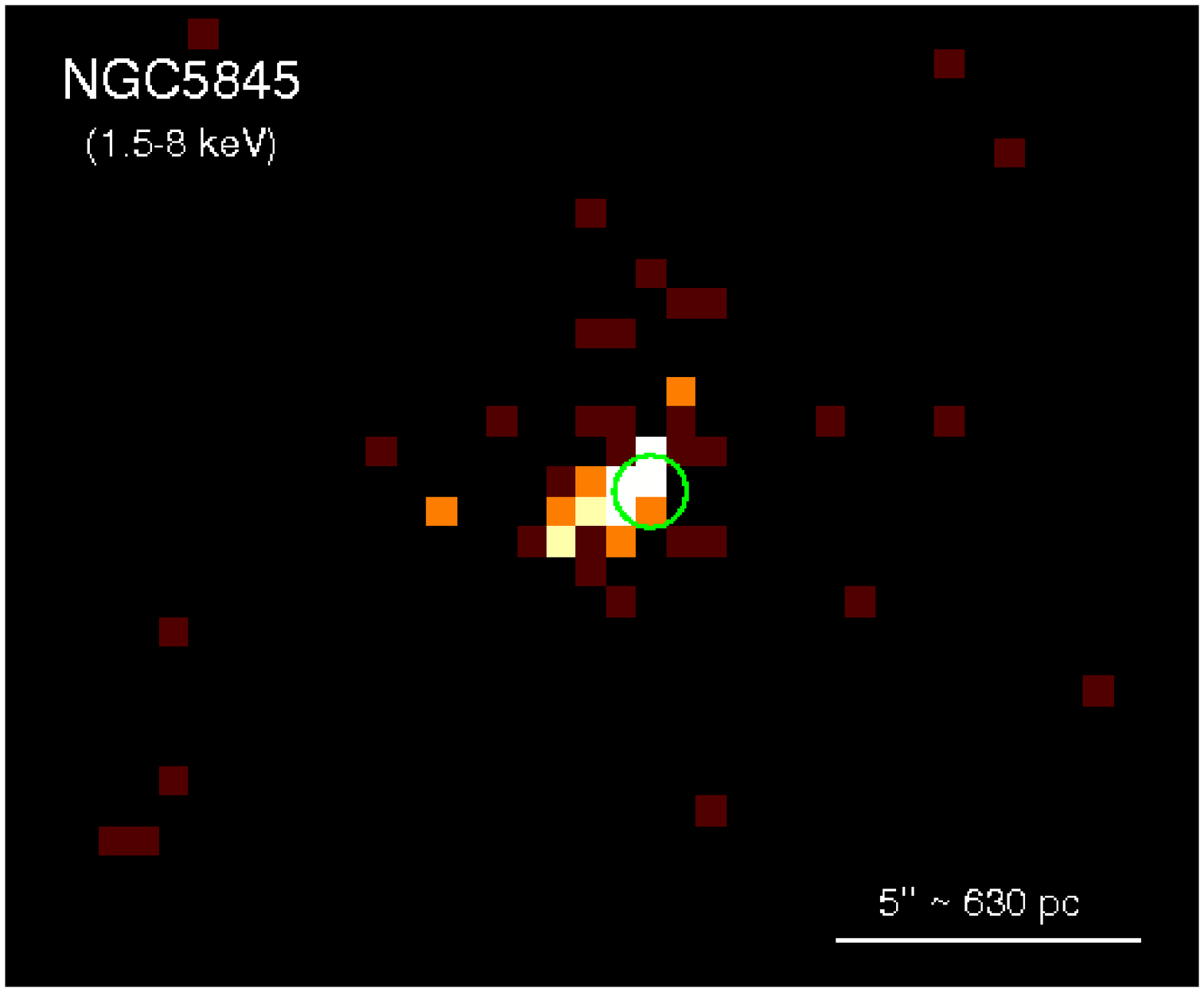}
\end{center}
%\plotone{acis4.ps}
\caption{As in Figure~1, but limited to the $1.5$--$8$ keV band. 
The green circles (radius of $0\farcs6$) mark the best 
available optical/IR position for the nuclei. North is up, 
East to the left.
\label{fig1b}}
\end{figure*}

\begin{figure}
%\epsscale{.80}
%\includegraphics[angle=270,scale=0.36]{colors_new4.ps}
\includegraphics[angle=270,scale=0.36]{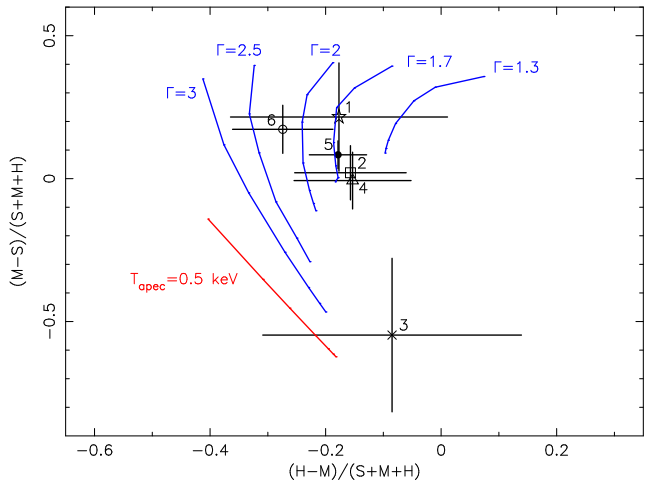}
%\plotone{color_test.ps}
\caption{X-ray color-color diagram for the nuclear emission 
in the six galaxies of our sample. We overplotted the expected 
colors for an optically-thin thermal plasma model 
and for power-law models of different photon indices, 
with varying line-of-sight column density ($n_{\rm H}$ 
running from $2 \times 10^{20}$ to $5 \times 10^{21}$ cm$^{-2}$ 
along each curve, from the bottom to the top).
The galaxies are labelled from 1 to 6 as in Table 1.
For NGC\,821, for which no point-like source 
is detected at the nuclear position, we plotted the average colors 
of the elongated, jet-like feature.
For NGC\,4697, we only used the 2003--2004 data, 
to minimize the effect of color changes due to the 
sensitivity degradation. The model curves are the colors 
expected for an observation in mid-2003.
\label{fig2}}
\end{figure}

\begin{figure}
%\epsscale{1.0}
%\includegraphics[angle=270,scale=0.35]{ngc4697_nucl_spectrum.ps}
\includegraphics[angle=270,scale=0.35]{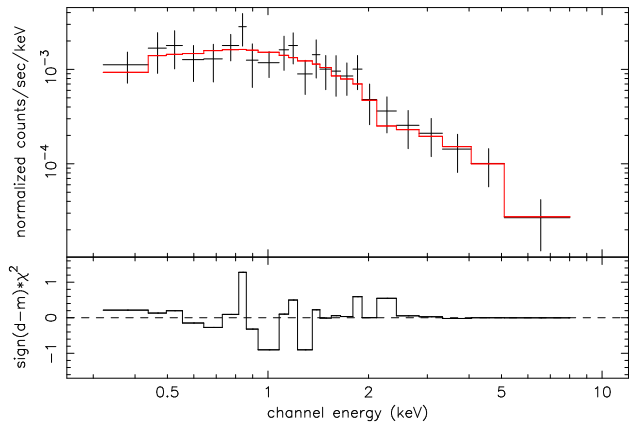}
%\plotone{ngc4697_diff.ps}
\caption{Spectrum of the nuclear source in NGC\,4697, 
with best-fit model and $\chi^2$ residuals. The data 
have been fitted with an absorbed power law, of photon 
index $\Gamma = 1.5 \pm 0.2$ (Table 3). 
\label{fig3}}
\end{figure}

%%%\clearpage
\begin{table*}
\begin{center}
\caption{Spectral parameters and emitted luminosities from the nuclear
sources. All spectra have been fitted with an absorbed power law, 
and column density fixed to the line-of-sight galactic value. For each 
source, the first fit is based on the Cash statistics, the second (when
applicable) on the $\chi^2$ statistics. Errors are 90\% 
confidence levels for one interesting parameter ($\Delta \chi^2 = 2.7$)
\label{tbl-3}}
\begin{tabular}{l@{\ \ \ }c@{\ \ \ }c@{\ \ \ }c@{\ \ \ }c@{\ \ \ }r}
%\begin{deluxetable}{lccc}
%\tabletypesize{\scriptsize}
%\rotate
%\tablecaption{Sample table taken from \citet{treu03}\label{tbl-1}}
%\tablewidth{0pt}
%\tablehead{
\tableline\tableline
Galaxy & Galactic $n_{\rm H}$\tablenotemark{a} 
&  Photon index $\Gamma$ & Normalization
at 1 keV & $\chi^2$/dof
& $L_{0.3-10 {\rm keV}}$\tablenotemark{b}\\
 & ($10^{20}$ cm$^{-2}$) & & ($10^{-6}$ ph. keV$^{-1}$ cm$^{-2}$ s$^{-1}$) 
 & &($10^{38}$ erg s$^{-1}$)\\
\tableline\\[-3pt]
NGC\,821 (``jet'') & $6.4$ & $1.53_{-0.31}^{+0.32}$& $5.3_{-1.0}^{+1.1}$ 
	&& $25.5 \pm 2.3$\\[4pt]
	& &$1.50_{-0.29}^{+0.31}$ & $4.9 \pm1.0$ & $7.5/6$ 
	& $24.4 \pm 2.2$\\[6pt]
NGC\,3377 & $2.9$ & $1.65^{+0.32}_{-0.25}$ & $3.3^{+0.6}_{-0.6}$ 
	&& $3.6^{+0.7}_{-0.8}$\\[4pt]
	& &$1.85^{+1.6}_{-1.2}$ & $3.9^{+1.8}_{-1.8}$ & $1.6/7$
	& $3.4^{+2.6}_{-1.7}$ \\[6pt]
NGC\,4486B\tablenotemark{c} 
	& $2.6$ & $2.4^{+0.5}_{-0.5}$ & $1.6^{+0.5}_{-0.5}$ &
	& $2.6^{+1.3}_{-1.0}$ \\[4pt]
%	&  & \multicolumn{4}{c}{Not applicable}  \\[6pt]
NGC\,4564 & $2.4$ & $2.4^{+0.4}_{-0.4}$ & $4.5^{+1.1}_{-1.2}$ &
	& $5.9^{+1.2}_{-1.8}$\\[4pt]
	&  & $1.6^{+1.5}_{-2.1}$ & $5.3^{+3.7}_{-3.7}$ & $0.76/4$
	& $10.5^{+25.3}_{-5.9}$ \\[6pt]
NGC\,4697\tablenotemark{d} & $2.1$ & $1.53^{+0.23}_{-0.22}$ & $2.9^{+0.5}_{-0.5}$ &$6.6/22$
	& $3.9^{+0.2}_{-0.2}$\\[6pt]
%	&  & $1.39^{+0.33}_{-0.31}$ & $3.1^{+0.6}_{-0.6}$ & $10.7/7$
%	& $4.8^{+1.2}_{-0.9}$& \\[6pt]
NGC\,5845 &$4.3$ & $2.03^{+0.24}_{-0.29}$ & $5.4^{+0.5}_{-1.6}$ &
	& $24.0^{+2.5}_{-2.5}$ \\[4pt]
	& & $1.4^{+0.8}_{-0.8}$ & $5.1^{+2.1}_{-2.2}$ & $1.3/7$
	& $39^{+19}_{-25}$\\[5pt]
\tableline
\end{tabular}
\tablenotetext{a}{from Dickey \& Lockman 1990.}
\tablenotetext{b}{Emitted luminosity inferred from the spectral fit.}
\tablenotetext{c}{Fitted with Cash statistics only.} 
%using WebPimms with a fixed photon index $\Gamma = 1.7$}
%\tablenotetext{c}{An emitted luminosity $\approx 4.3 \times 10^{38}$
%(scaled to the distance adopted here) was obtained by Sarazin et
%al.~(2001), assuming a bremsstrahlung spectrum with $kT = 8.1$ keV} 
%\tablenotetext{b}{Yet another sample footnote for table~\ref{tbl-2}}
\tablenotetext{d}{Fitted with the $\chi^2$ statistics.}
%\tablecomments{We can also attach a long-ish paragraph of explanatory
%material to a table.}
\end{center}
\end{table*}

%%%\clearpage

\subsection{Scattering and obscuration of the nuclear SMBH?}

The nuclear X-ray source in NGC\,5845 appears extended 
on the scale of a few hundred pc, larger 
than the FWHM of the {\it Chandra}/ACIS PSF for 
a point source (Section 3.1). The count rate 
is insufficient to map the X-ray colors or 
the spectral properties across the nuclear region.
However, we can at least say that the ``fuzzy'' 
emission around the nuclear position is not 
significantly softer than a typical AGN-like 
power-law spectrum: the emission is extended 
even in the hard band ($2$--$10$ keV).  
The simplest explanation is that the emission is due 
to a few faint, unresolved LMXBs in the inner region.
If that was the case, the true luminosity 
of the SMBH in NGC\,5845 would be 
a few times lower than estimated 
in Table 3.

Alternatively, we note that the X-ray morphology is reminiscent
of the extended X-ray emission associated with 
obscured Seyfert nuclei (for NGC\,4151: 
Elvis, Briel \& Henry~1983; Elvis et al.~1990;
Ogle et al.~2000; for NGC\,1068: Ogle et al.~2003), 
which is understood as scattered nuclear emission. 
Dust is known to be present in the nuclear 
region of most ellipticals, especially 
those with disky isophotes (van den Bosch et al.~1994; 
Kormendy et al.\ 2005 and references therein).
The nuclear region of NGC\,5845 contains a dusty disk of radius 
$\approx 100$ pc (Quillen, Bower \& Stritzinger 2000), 
i.e. an order of magnitude larger than the Bondi 
accretion radius.
In fact, NGC\,5845 is one of only two galaxies 
known for containing both a dust disk 
and an associated, nuclear stellar disk 
(the other one is NGC\,4486A). It has been 
suggested (Kormendy et al.\ 2005; Kormendy et al.\ 1994)
that, as cold gas and dust settle 
towards the center, they form stars and build 
a stellar disk.

In NGC\,5845, the disk is seen almost edge-on (Figure 5).
Thus, it is possible 
that most of the direct X-ray emission from the SMBH 
along our line of sight is obscured by dust, 
and most of the observed, extended emission comes 
from the surrounding photo-ionized plasma which is 
scattering the nuclear continuum.
If this was the case, the true SMBH luminosity 
in NGC\,5845 could instead be at least an order of magnitude 
higher than estimated in Table 3.
%Assuming that the bolometric luminosity initially 
%emitted (mostly in the optical/UV/X-ray bands)   
%was $\approx 0.1 \dot{M}_{\rm B} c^2 \approx 6 \times 10^{41}$ 
%erg s$^{-1}$, we expect a fraction $0.1 g \dot{M}_{\rm B} c^2$ 
%to be intercepted by the obscuring disk/torus and re-radiated 
%in the far-infrared. 
Given the apparent thin-disk geometry 
of the dusty feature (Figure~5), we may expect 
that the disk/torus can intercept 
and re-radiate in the far-infrared 
a fraction $\sim 10$--$20\%$ of the bolometric luminosity. 
%$g\sim 0.1$--$0.2$, 
If the far-infrared nuclear luminosity 
$\ga 10^{41}$ erg s$^{-1}$, it is potentially 
detectable by {\it Spitzer} (flux density $\sim 10$ mJy 
in the {\it Spitzer}/MIPS bands). This will 
be investigated in a follow-up work, for which {\it Spitzer} 
time has already been obtained.
For the present paper, we assume that 
the observed X-ray luminosity of the SMBH 
in NGC\,5845 is a good approximation 
of the emitted luminosity.

A dusty disk or torus of outer radius $\approx 200$ pc 
is also apparent in the nuclear region of NGC\,4697 (Figure 6). 
However, obscuration or scattering of the central X-ray 
source may not be as significant as in NGC\,5845 
because our viewing angle is $\approx 75^{\circ}$.
A more detailed investigation of the physical 
processes at work in a dusty nuclear disk is beyond 
the scope of this paper.

%%%\clearpage
\begin{figure}[t]
%\epsscale{1.15}
%\includegraphics[angle=0,scale=0.45]{ngc5845_hst.ps}
\includegraphics[angle=0,scale=0.45]{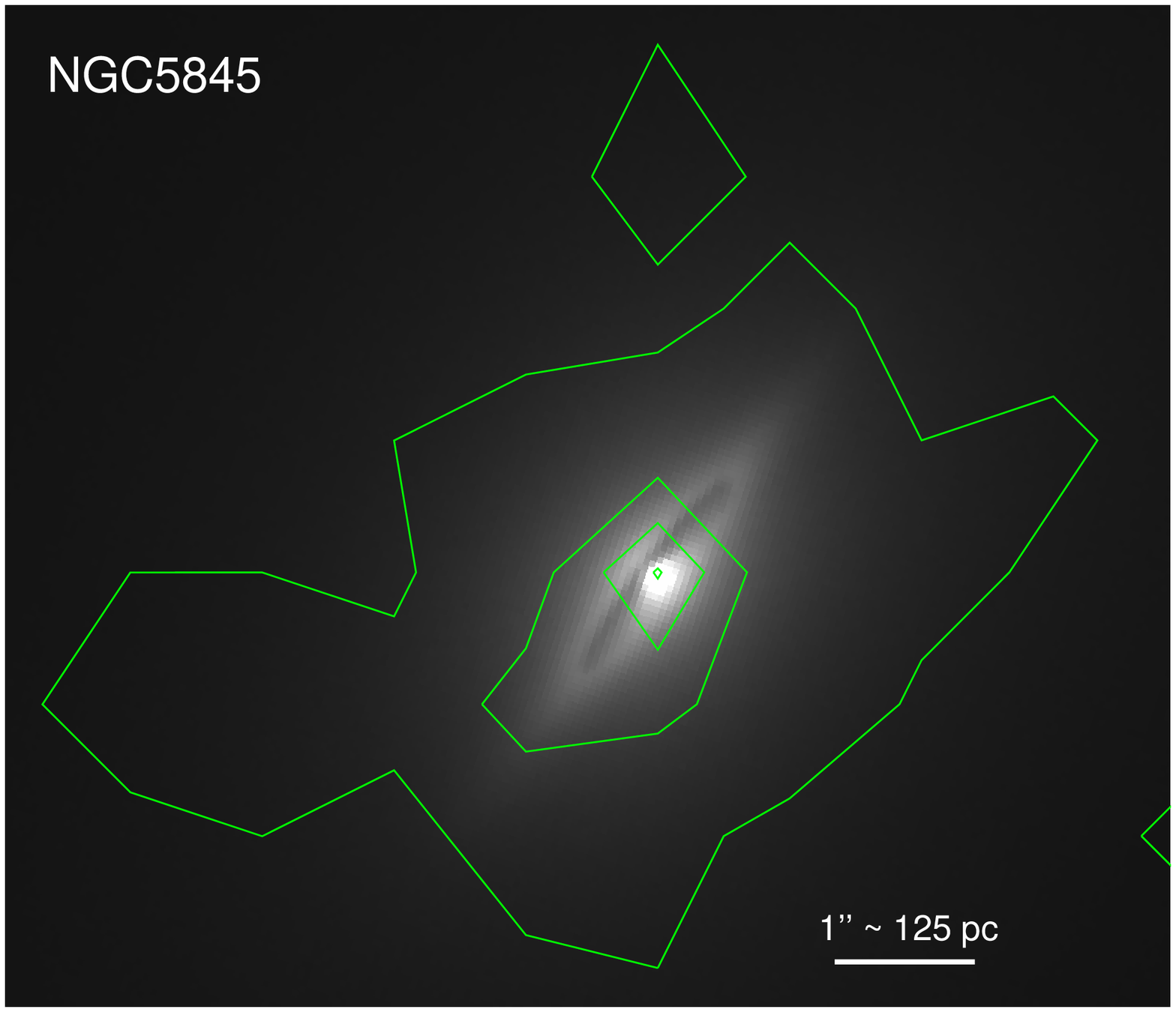}
%\includegraphics[angle=0,scale=0.47]{ngc5845_isophotes.ps}
%\plotone{ngc4697_color_crop.eps}
\caption{Greyscale {\it HST}/WFPC2 image (F555W filter) 
of the inner region of NGC\,5845, with X-ray contours 
of the nuclear emission overplotted (cf. Figure 1, 
bottom right panel). North is up, East to the left.
The nucleus contains an 
$\approx 100$-pc-radius dusty and stellar disk, 
which may contribute to obscuring or scattering 
the central X-ray source. In an alternative scenario, 
the dusty disk may reduce the accretion rate onto the BH 
by storing cold gas and triggering star formation.
\label{fig13}}
\end{figure}

\begin{figure}[t]
%\epsscale{1.15}
%\includegraphics[angle=0,scale=0.45]{ngc4697_hst2.ps}\\
%\includegraphics[angle=0,scale=0.45]{ngc4697_hst1.ps}
\includegraphics[angle=0,scale=0.45]{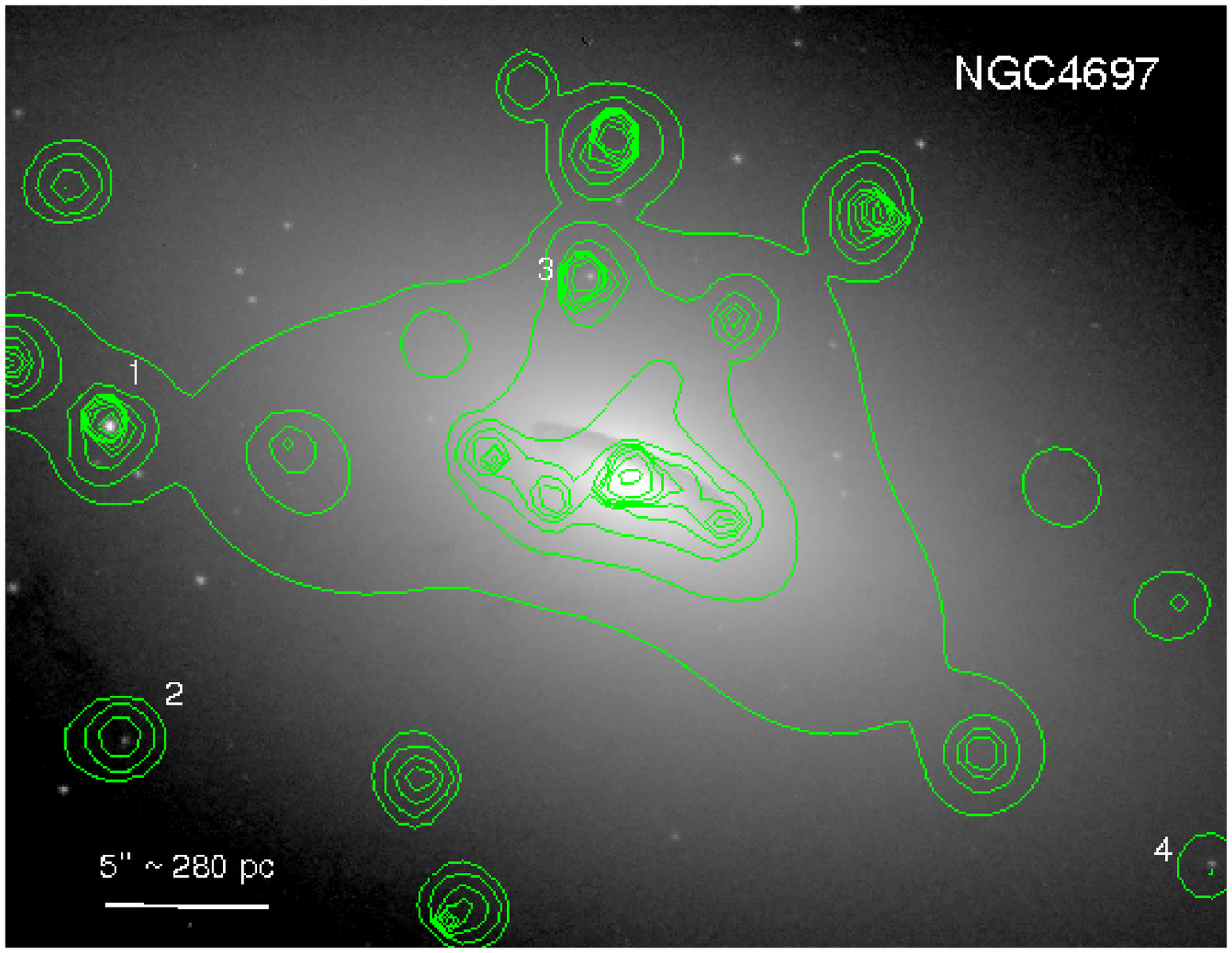}\\
\includegraphics[angle=0,scale=0.45]{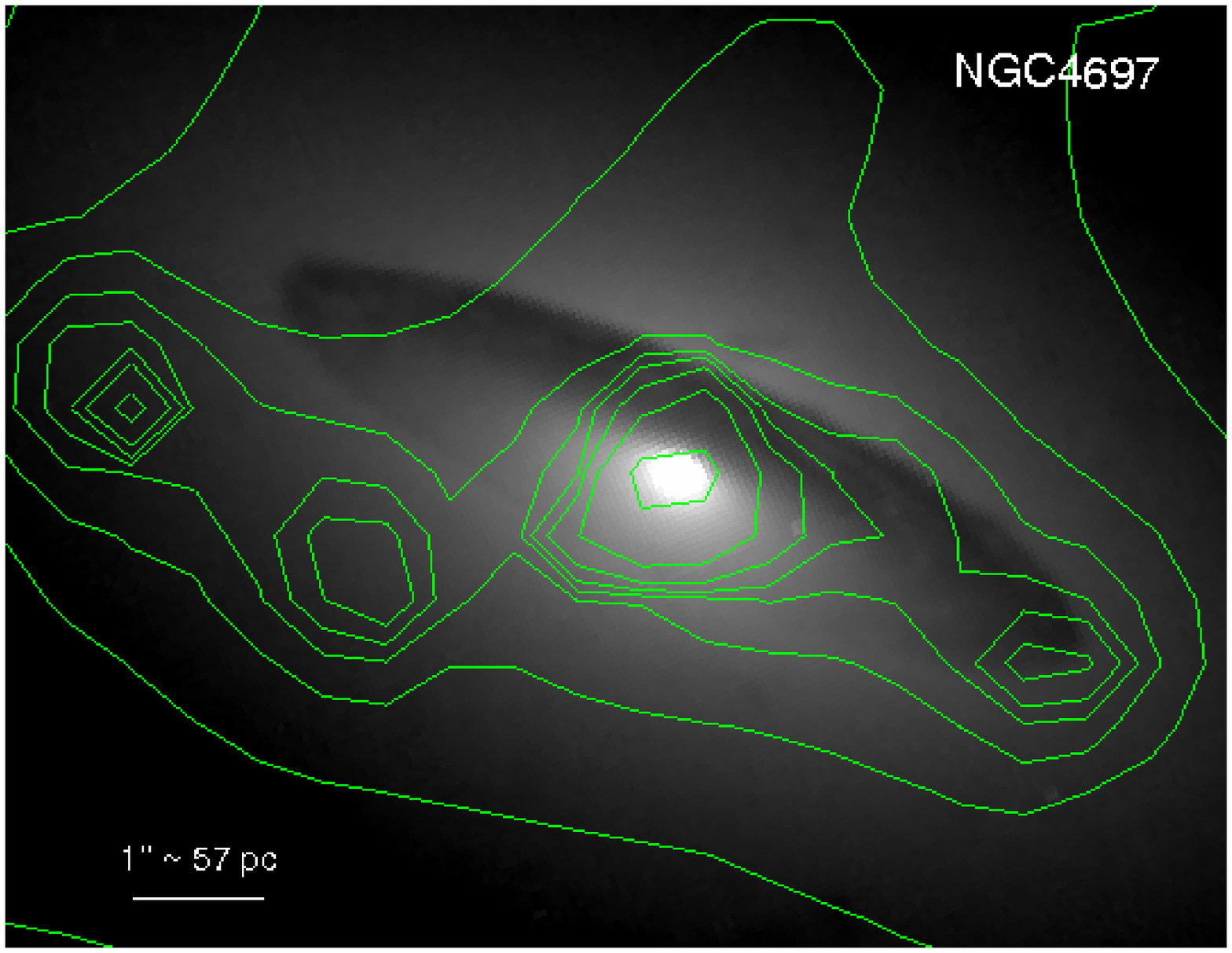}
%\includegraphics[angle=0,scale=0.47]{ngc5845_isophotes.ps}
%\plotone{ngc4697_color_crop.eps}
\caption{Greyscale {\it HST}/ACS image (F475W filter) 
of the inner region of NGC\,4697 (top panel), 
with a close-up view of the nuclear region (bottom panel). 
In both panels, North is up, East to the left.
The X-ray contours are on a square-root scale with arbitrary 
zeropoint. 
We improved the relative astrometry of the {\it HST} 
and {\it Chandra} images 
by taking advantage of a few X-ray sources 
with an optical counterpart (possibly globular clusters 
in NGC\,4697). Four of them are visible here 
(in addition to the nuclear source), labelled 
as ``1'' through ``4''. The optical nucleus 
is point-like in {\it HST} (size $\la 0\farcs1 \approx 5$ pc) 
and is probably a stellar cluster.
\label{fig14}}
\end{figure}
%%%\clearpage

%(e.g., {\footnotesize NGC 4151 - Elvis et al 1983, ApJ, 268, 105 
%and Ogle et al 2000, ApJ, 545, L810; Elvis et al 1990, ApJ, 361,
%459}), 

\subsection{Jet-like features}

The morphologies of the X-ray emission 
in the nuclear regions suggest possible 
jet-like features in two of our target galaxies, 
NGC\,821 and NGC\,3377. The former was discussed 
in Fabbiano et al.~(2004), who suggest 
that the X-ray flux is consistent with synchrotron
emission in a jet, or with hot thermal plasma 
shocked by intermittent nuclear outbursts.
For NGC\,3377, we compared the {\it Chandra} image 
with an archival {\it HST}/WFPC2 image 
in the F555W filter, 
to determine whether the elongated feature 
in the X-ray image has an optical counterpart; however, 
we do not see any obvious distortions in the optical 
isophotes (Figure 7, top panel). 

We then searched for possible fainter 
optical signatures in the nuclear region of NGC\,3377.
We determined the mean 2-D surface brightness distribution
of the central region of NGC\,3377 with the help of a boxcar
pixel-smoothing technique and an isophotal fitting routine
written in {\footnotesize IRAF}. Both techniques were recently employed
to uncover spiral and bar structures in seemingly normal dwarf
elliptical galaxies in the Virgo (Jerjen, Kalnajs \& Binggeli~2000; 
Barazza, Binggeli \& Jerjen~2002) and Coma clusters (Graham, 
Jerjen \& Guzm\'{a}n~2003).
The reconstructed model galaxy is subtracted from
the original, leaving just the small-scale substructures
in the residual image. The result for NGC\,3377 derived in
this manner is shown in the bottom panel of Figure 7: 
we find a small, brighter stellar disk  
but no feature corresponding to the X-ray elongation, 
neither parallel not perpendicular to the disk plane.

None of the six galaxies in our sample have 
significant radio detections; however, five have
at least reliable upper limits to their radio-core emission, 
from VLA observations.
In NGC\,821, the radio-core flux at 5 GHz is $< 0.3$ mJy 
(Wrobel \& Heeschen 1991); in NGC\,3377, NGC\,4697 and NGC\,5845, 
the radio flux at 8.4 GHz is $< 0.1$ mJy 
(Krajnovic \& Jaffe 2002). In NGC\,4564, the upper 
limit is $0.52$ mJy at 8.4 GHz (Wrobel \& Herrnstein 2000); 
however, the core is marginally detected 
at the level of $\approx 2$ mJy at 1.4 GHz, 
in the NRAO/VLA Sky Survey. A comparison of radio 
and X-ray nuclear fluxes is discussed in Paper II.

%%%\clearpage
\begin{figure}
%\epsscale{1.0}
\includegraphics[angle=0,scale=0.45]{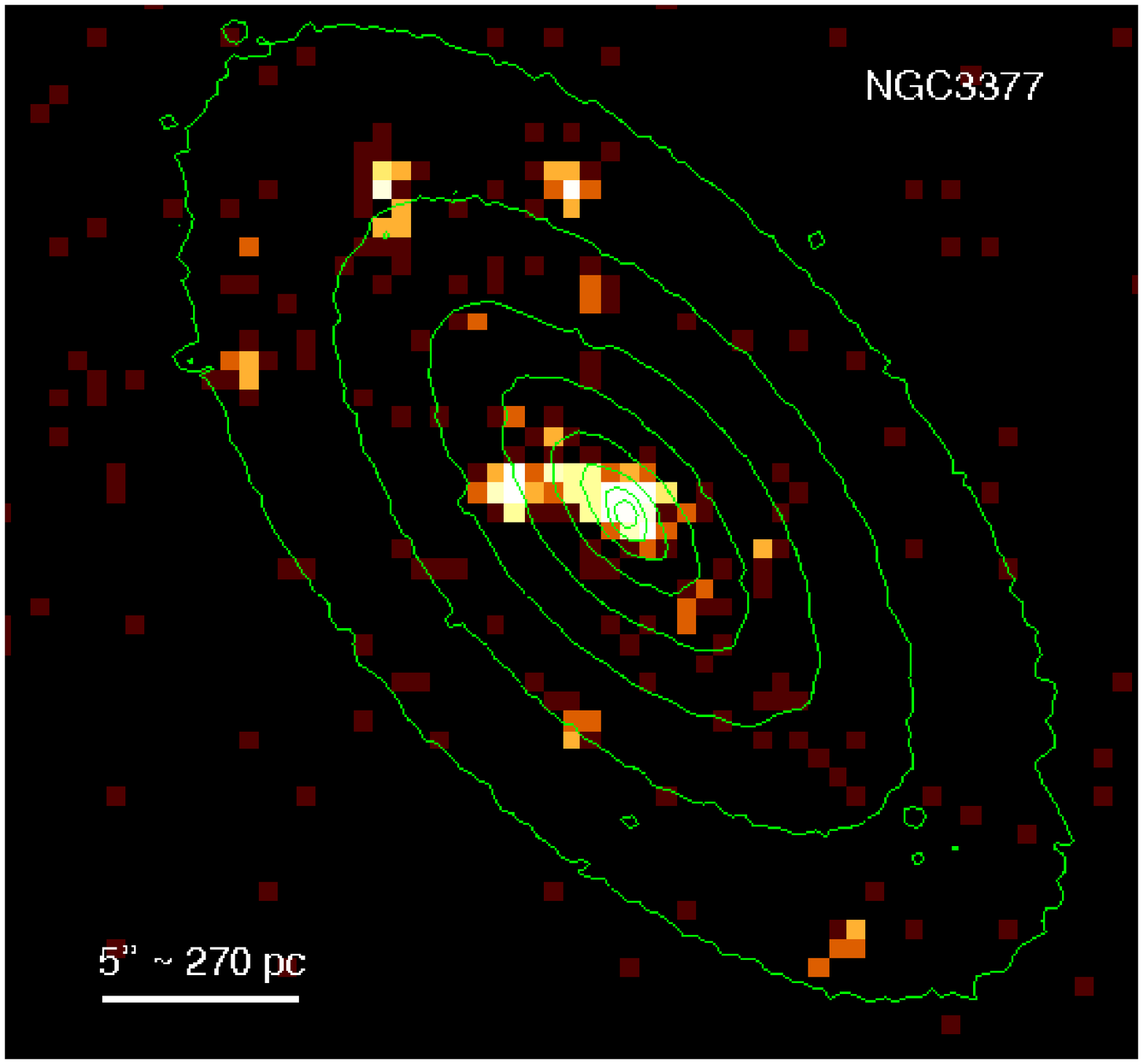}\\
\includegraphics[angle=0,scale=0.45]{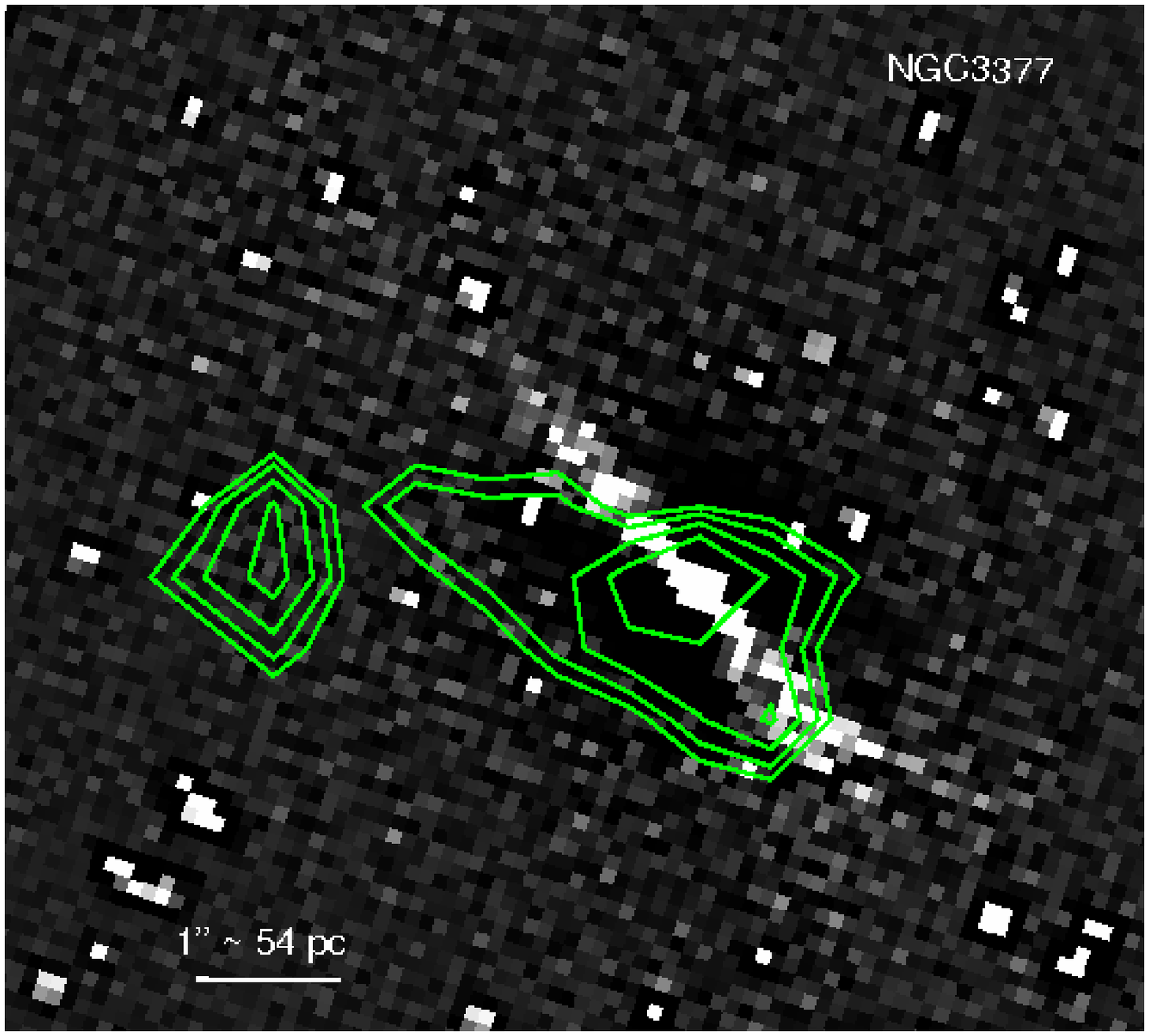}
%\includegraphics[angle=0,scale=0.45]{ngc3377_new.ps}\\
%\includegraphics[angle=0,scale=0.45]{ngc3377_residual_3.ps}
%\plotone{surf_bright.ps}
\caption{Comparison between X-ray ({\it Chandra}/ACIS, 
$0.3$--$8$ keV band) 
and optical ({\it HST}/WFPC2, in the F555W filter) 
images of the inner region of NGC\,3377. 
Top panel: the elongated X-ray emission feature 
at the nuclear position does not have any correspondence 
in the optical isophotes. 
Bottom panel: X-ray emission contours overplotted 
onto a greyscale residual optical image. 
The residual image was obtained by subtracting 
the mean 2-D surface brightness from the original 
optical image (see text for details). A bright 
stellar disk is found in the nuclear region, but has 
no apparent association with the elongated X-ray feature.
In both panels, North is up, East to the left.
\label{fig12}}
\end{figure}
%%%\clearpage

\section{X-ray emitting, diffuse gas}

\subsection{X-ray brightness profiles}

Unresolved X-ray emission in elliptical galaxies 
is a combination of faint low-mass X-ray binaries (LMXBs)
and hot gas, with $kT \sim 0.3$--$0.7$ keV 
(Trinchieri \& Fabbiano 1985; Blanton, Sarazin \& Irwin 2001;
Sivakoff, Sarazin \& Irwin 2003; Randall, Sarazin \& Irwin 2004).
In this work, we focus on the nuclear BH activity 
and do not set out to study the X-ray source population 
and large-scale structure of the galaxies. However, 
we still need to investigate the diffuse hot gas component, 
because it provides a constraint to the energy budget 
of the nuclear BH. From the surface brightness 
and temperature of the diffuse emission, we can in principle 
estimate the density of the hot gas in the nuclear region; 
hence, we can obtain a lower limit to the gas density 
available for accretion onto the BH, which can be 
compared to the observed luminosity, to constrain 
the accretion efficiency.

To obtain the surface brightness of the unresolved 
emission (and, in particular, the thermal gas component), 
firstly we filtered the event files 
to remove the point sources detected with {\tt {wavdetect}}.
In the case of NGC\,4697, we combined all four observations 
to reach a deeper detection limit for the point sources; 
then we analysed the unresolved emission separately 
for the 2000 and 2003--2004 datasets.
For each galaxy, we created a stack of concentric annular regions 
extending up to $120\arcsec$ from the nucleus, and a background 
annulus outside this region. We used the {\footnotesize {CIAO}} 
task {\tt {dmextract}} to obtain the background-subtracted 
radial profiles. Since we are only interested here 
in the diffuse gas contribution, we restricted the energy 
range to $0.3$--$1.5$ keV, where the hot thermal plasma 
component is relatively stronger, compared with 
the unresolved LMXB contributions. Above 
$\approx 1.5$ keV, the hot gas component is, instead, 
negligible, and the unresolved emission is dominated by the 
power-law LMXB component. 
 
To determine the physical properties of the extended 
X-ray emission, we need to measure the radial profiles 
of the gas surface brightness $S(r)$, and the central 
surface brightness $S_0$ (a distance-invariant quantity). 
The surface brightness $S(r)$ does not come directly 
from the observations: what we have instead is 
the surface density $S'(r)$ of the count rates 
in the $0.3$-$1.5$ kev band, for the total 
unresolved emission, including contributions 
from both the interstellar gas and faint LMXBs.
We show $S'(r)$ for five of the six galaxies 
in our sample (Figures~8, 9). The strongest 
emission is found in NGC\,4697 (Figure~10; see also 
Sarazin et al.~2001). On the other end, 
no unresolved emission is detected in NGC\,4486B, 
though the relatively higher background level of that observation 
makes it more difficult to obtain a tight upper limit.
To calculate the density of the hot gas, 
we need to convert the observed count rates into 
physical flux units, and to estimate what fraction 
of the unresolved emission is due to diffuse gas.
This requires a spectral analysis.

Only two galaxies (NGC\,4697 and NGC\,3377) have enough detected 
counts in the unresolved X-ray emission to allow 
for spectral fitting. After excluding the discrete 
sources found by {\tt{wavdetect}}, we used the 
{\footnotesize{CIAO}} script {\tt{acisspec}} to extract 
the spectrum of the unresolved emission from a $90\arcsec$ 
circle, taking an annulus between $120\arcsec$ and 
$150\arcsec$ as background region. For NGC\,4697, 
we extracted separate spectra (with their respective 
response files) for the 2000 and 2003--2004 observations; 
we then combined the two spectra using the algorithm 
of Page et al.~(2003).
We fitted the spectra in {\footnotesize{XSPEC}}, 
assuming an absorbed power-law plus a single-temperature 
thermal plasma model, {\tt{wabs*(vapec + po)}}.
First, we fixed the absorption column density to the Galactic value, 
the metal abundance of all elements to the solar value, 
and the photon index of the power-law component 
to $1.7$ (average value for LMXBs), to reduce 
the number of free parameters (Tables 4, 5).
This already produces a good fit ($\chi^2 = 43.0/51$)
As previously noted ({\it ROSAT} study: Pellegrini 
\& Fabbiano 1994; {\it Chandra} study: Sarazin et al.~2001), 
the interstellar gas in NGC\,4697 is cooler ($kT \approx 0.3$ keV) 
than is usually found in X-ray bright ellipticals; 
this may be the consequence of a shallow gravitational potential. 
We further improved the fit by varying the metal abundances 
of the main elements: we find that an overabundance 
of C and N over the $\alpha$ elements 
by a factor of $\sim 5$--$10$ significantly 
reduces the $\chi^2$ below 1 keV. No significant Mg or Si lines 
are detected (Figure 11), an indication that the X-ray emitting gas 
may not be strongly metal-enriched from Type-II SNe.
The gas in NGC\,3377 is consistent with a higher temperature 
($kT \approx 0.6$ keV), 
although the uncertainty is larger, due to the limited 
number of detected counts (only $\approx 130$; Figure 12). 
There are not enough counts to vary the metal abundances.

By fitting a spectral model to the unresolved X-ray emission, 
we can estimate the conversion factors between 
the observed count-rate density $S'(r)$ 
(units of counts s$^{-1}$ arcsec$^{-2}$ 
in a specific band), and the bolometric surface brightness $S(r)$  
of the diffuse hot gas. 
For the spectral model fitted 
to NGC\,3377, we obtain a conversion factor of 
$1$ count s$^{-1}$ arcsec$^{-2}$ in the $0.3$--$1.5$ keV band
$= (2.5 \pm 0.5)$ erg cm$^{-2}$ s$^{-1}$ arcsec$^{-2}$;  
$\approx 50\%$ of the unresolved counts in the $0.3$--$1.5$ keV band 
are from truly diffuse emission. This is an average value 
over the whole extraction region; in fact, 
the diffuse emission may be more centrally peaked 
(Kim \& Fabbiano 2003).
For the 2000 observation of NGC\,4697, 
$1$ count s$^{-1}$ arcsec$^{-2}$ $= (4.8\pm 0.4)$ erg cm$^{-2}$ 
s$^{-1}$ arcsec$^{-2}$; in 2003--2004, $1$ count s$^{-1}$ 
arcsec$^{-2}$ $= (6.7\pm 0.5)$ erg cm$^{-2}$ 
s$^{-1}$ arcsec$^{-2}$ (this difference is due to the 
decrease in sensitivity of the detector); 
$\approx 80\%$ of the unresolved counts are from hot gas.
The total emitted luminosity from the hot gas is $\approx 10^{38}$ 
erg s$^{-1}$ in NGC\,3377 and $\approx 10^{39}$ 
erg s$^{-1}$ in NGC\,4697.

We can use these values to estimate the conversion 
factors for the other galaxies in our sample, 
for which we cannot do any spectral analysis. 
We can safely assume that the temperature 
of the hot ISM is in the range $\sim 0.3$--$0.7$ keV.
The relative contribution of hot gas and LMXB emission 
varies from galaxy to galaxy (see a discussion 
in Kim \& Fabbiano 2003); our choice of a soft 
energy band ($0.3$--$1.5$ keV) should reduce 
the uncertainty caused by LMXB contamination. 
%however, all our target 
%galaxies have been selected as X-ray faint 
%ellipticals, for which the hot ISM 
%luminosity is typically $\sim 10\%$ 
%of the total LMXB luminosity (**ADD REFS**). 
The conversion factor between unresolved count rate  
and flux of the truly diffuse emission 
depends strongly on the detection threshold 
of the point sources: we used NGC\,4697 to estimate this effect.
When the point-source detection limit is $1 \times 10^{37}$ erg s$^{-1}$,
we find that $\approx 80\%$ of the unresolved counts 
in the $0.3$--$1.5$ keV band are 
from the ISM, as mentioned earlier; 
%(averaged over the $90\arcsec$ extraction region); 
$\approx 50\%$ for a detection limit of $1 \times 10^{38}$ 
erg s$^{-1}$; and only $\approx 35\%$ for a detection 
limit of $3 \times 10^{38}$ erg s$^{-1}$. 
%These fraction 
%are slightly higher in the central regions, because 
%the diffuse emission is more concentrated than the LMXBs  
%(see also Kim \& Fabbiano 2003).
In conclusion, taking into account these arguments, 
we adopt a conversion factor between count-rate surface 
density and bolometric surface brightness of 
$1$ count s$^{-1}$ arcsec$^{-2}$ in the $0.3$--$1.5$ keV band
$= (3.0 \pm 1.5)$ erg cm$^{-2}$ s$^{-1}$ arcsec$^{-2}$ 
from the hot gas, for NGC\,4564 and NGC\,5845.

%%%\clearpage
\begin{figure}
%\epsscale{1.0}
\includegraphics[angle=270,scale=0.36]{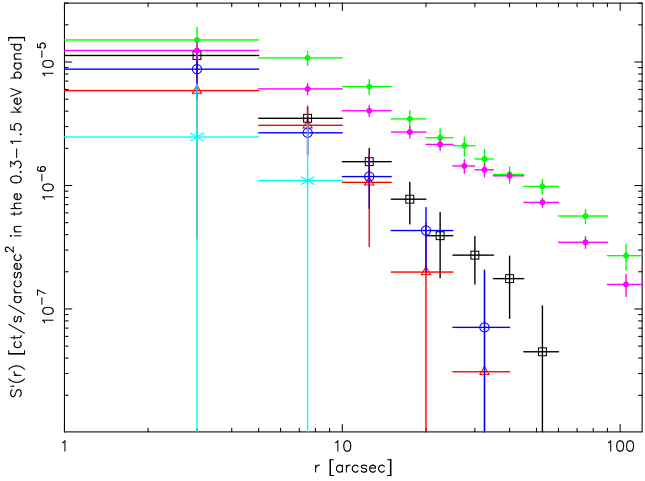}
%\includegraphics[angle=270,scale=0.36]{rprofiles_final1.ps}
%\plotone{surf_bright.ps}
\caption{Count-rate surface density profiles of the unresolved emission 
in the $0.3$--$1.5$ keV band (units of ACIS-S counts 
s$^{-1}$ arcsec$^{-2}$), 
plotted as a function of angular distance on the sky. 
From the brightest to the faintest, symbols (and 
colors for the online version of the paper) 
are as follows: filled circles for NGC\,4697 (green 
for the 2000 observations, magenta 
for the 2003--2004 observations); open squares 
(black) for NGC\,3377; open circles (blue) 
for NGC\,5845; open triangles (red) 
for NGC\,4564; crosses (cyan) for NGC\,4486B.\label{fig4}}
\end{figure}

\begin{figure}
%\epsscale{1.0}
\includegraphics[angle=270,scale=0.36]{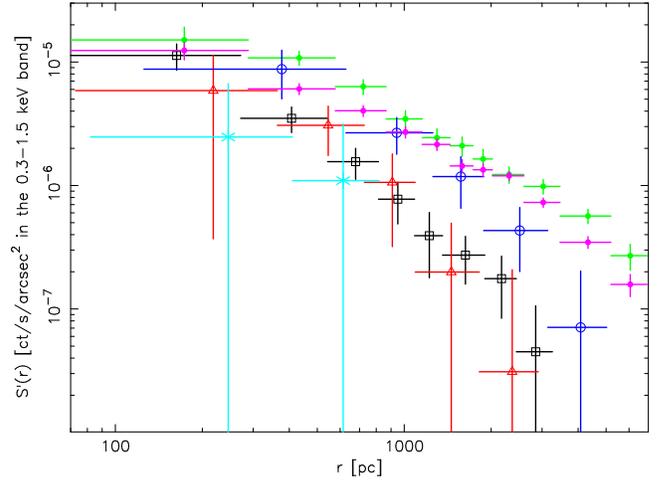}
%\includegraphics[angle=270,scale=0.36]{rprofiles_final1_pc.ps}
%\plotone{surf_bright.ps}
\caption{Count-rate surface density profiles of the unresolved emission 
in the $0.3$--$1.5$ keV band (units of ACIS-S counts 
s$^{-1}$ arcsec$^{-2}$), 
plotted as a function of projected physical distance from the nucleus. 
Symbols and colors are as in Figure 8.\label{fig5}}
\end{figure}

\begin{figure*}
\epsscale{1.15}
\includegraphics[angle=0,scale=0.48]{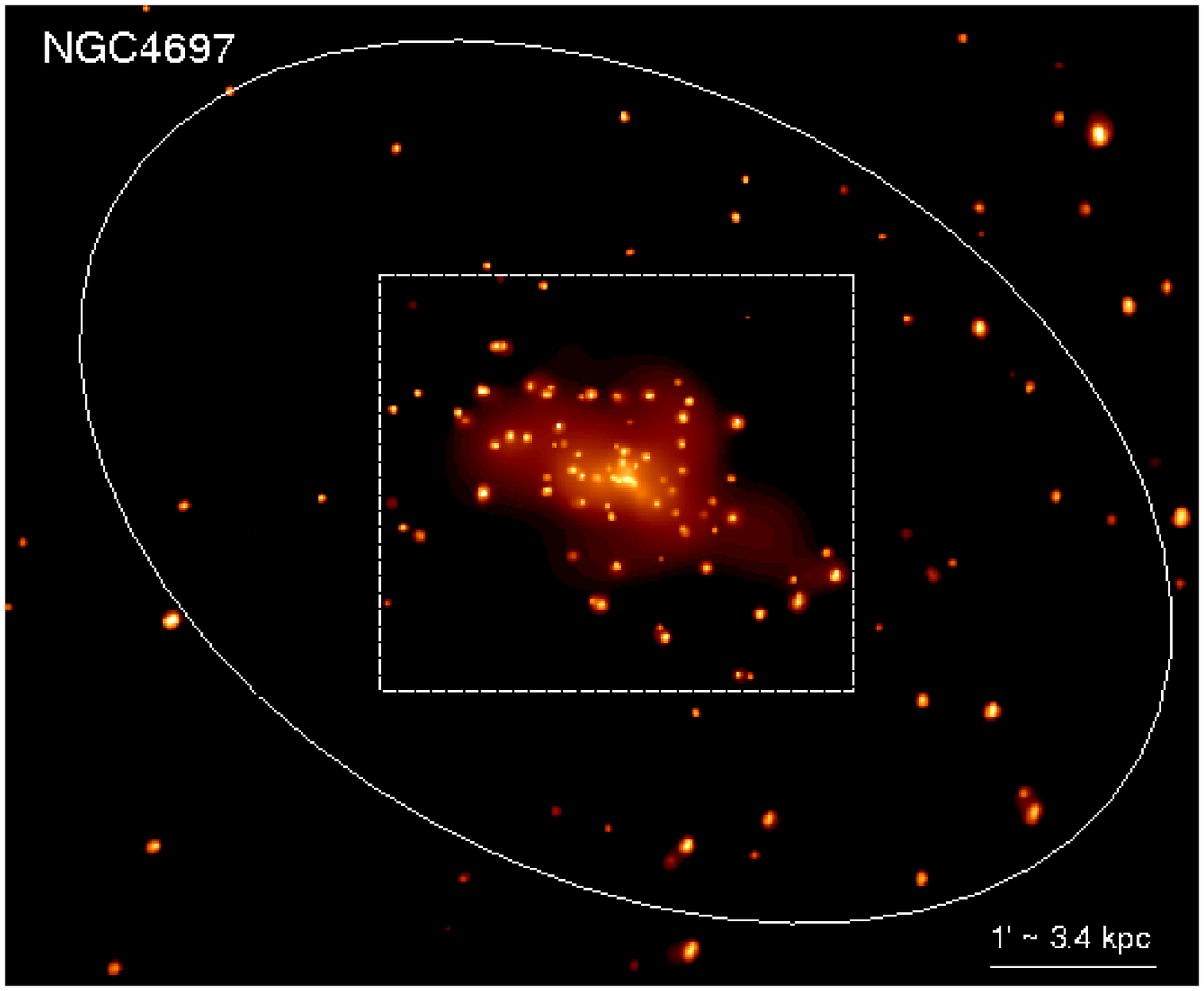}
\includegraphics[angle=0,scale=0.79]{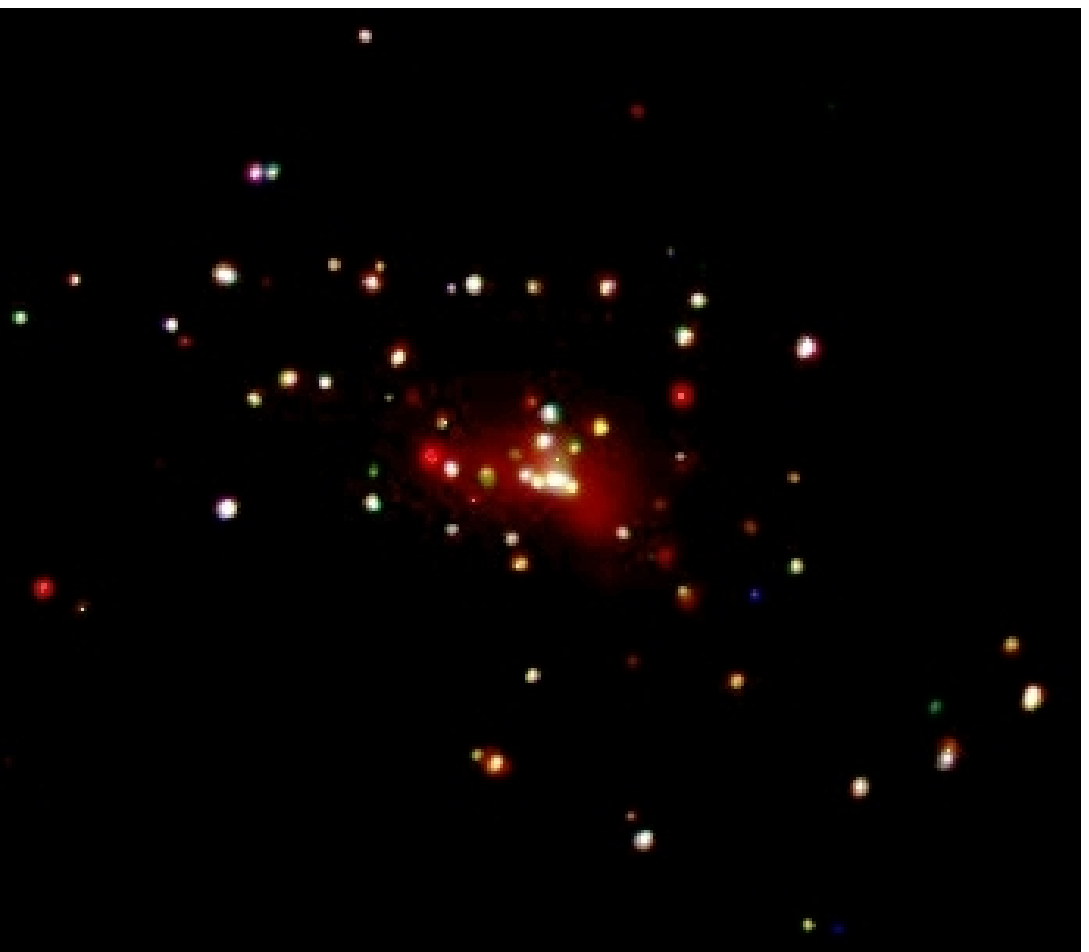}
%\includegraphics[angle=0,scale=0.48]{ngc4697_acis_box2.ps}
%\includegraphics[angle=0,scale=0.79]{ngc4697_color_new_crop.eps}
%\plotone{ngc4697_color_crop.eps}
\caption{Left panel: adaptively-smoothed ACIS-S image 
of NGC\,4697, in the $0.3$--$8$ keV band. The ellipse 
marks the $D_{25}$ isophote; the box (dashed white line) 
is shown in greater details on the right. Right panel: 
true-color X-ray image of the inner region of 
NGC\,4697. Red $= 0.3$--$1$ keV; 
green $= 1$--$2$ keV; blue $= 2$--$8$ keV.
In both panels, North is up, East to the left.\label{fig6}}
\end{figure*}

\begin{figure}
%\epsscale{1.0}
\includegraphics[angle=270,scale=0.35]{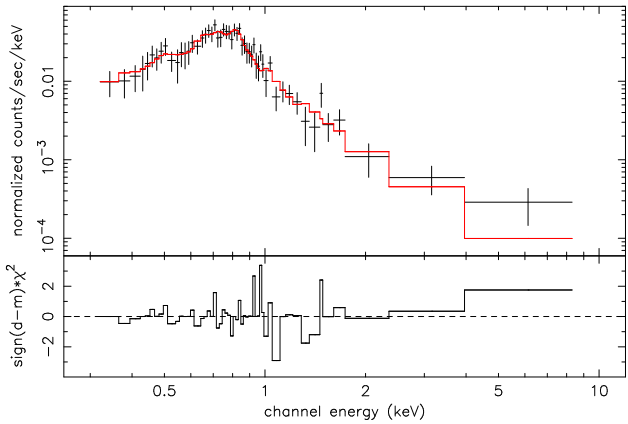}
%\includegraphics[angle=270,scale=0.35]{ngc4697_diff_spectrum.ps}
%\plotone{ngc4697_diff.ps}
\caption{Spectrum of the unresolved emission from NGC\,4697, 
with best-fit model and $\chi^2$ residuals. The data have been 
fitted with a variable-abundance thermal plasma model 
({\tt{vapec}} in {\footnotesize{XSPEC}}) plus a power-law. The fit parameters 
are listed in Table 4.\label{fig7}}
\end{figure}

\begin{figure}
%\epsscale{1.0}
\includegraphics[angle=270,scale=0.35]{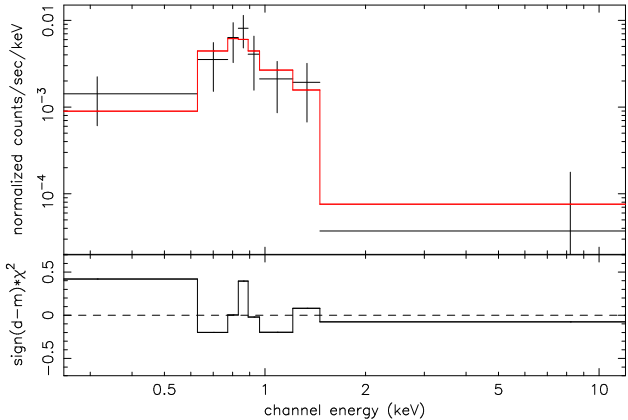}
%\includegraphics[angle=270,scale=0.35]{ngc3377_diff.ps}
%\plotone{ngc4697_diff.ps}
\caption{Spectrum of the unresolved emission from NGC\,3377, 
with best-fit model and $\chi^2$ residuals. The data have been 
fitted with a thermal plasma model 
({\tt{apec}} in {\footnotesize{XSPEC}}, 
with fixed solar abundance) plus a power-law. 
The fit parameters are listed in Table 5.\label{fig8}}
\end{figure}
%%%\clearpage

\begin{table}
\begin{center}
\caption{Best-fit parameters for the spectrum of the unresolved 
X-ray emission in NGC\,4697. Spectral model: 
{\tt wabs$_{\rm Gal}$*wabs*(vapec+po)}. Values in brackets were kept
fixed. Errors are 90\% 
confidence levels for one interesting parameter ($\Delta \chi^2 =
2.7$).
\label{tbl-5}}
\begin{tabular}{l@{\ \ \ \ \ }c@{\ \ \ \ \ }r}
\tableline\tableline\\[-6pt]
Parameter & Model 1 & Model 2\\
 & Value & Value\\
\tableline\\[-3pt]
$n_{\rm {H,Gal}}$\tablenotemark{a} & $(2.1)$ & $(2.1)$\\[6pt]
$n_{\rm {H}}$ & $(0.0)$ & $2.2^{+6.6}_{-2.2}$\\[6pt]
$\Gamma$\tablenotemark{b}  & $(1.7)$ & $(1.7)$\\[6pt]
$N_{\rm {pl}}$ at 1 keV\tablenotemark{c} &  $8.4^{+2.4}_{-2.2} \times 10^{-6}$
	& $8.1^{+2.3}_{-2.2} \times 10^{-6}$\\[6pt]
$kT_{\rm{gas}}$ (keV) &$0.34^{+0.02}_{-0.02}$ & $0.32^{+0.03}_{-0.03}$ \\[6pt]
$Z_{\rm C,N} (Z_{\odot})$ & $(1.0)$ & $9.1^{+7.3}_{-5.1}$\\[6pt]
$Z_{\rm 0-Ca} (Z_{\odot})$ & $(1.0)$ & $(1.0)$\\[6pt]
$Z_{\rm Fe,Ni} (Z_{\odot})$ & $(1.0)$ & $1.6^{+1.0}_{-0.6}$\\[6pt]
$K_{\rm{vapec}}$\tablenotemark{d} &$3.2^{+0.3}_{-0.4} \times 10^{-5}$ 
	 & $2.7^{+1.3}_{-0.9} \times 10^{-5}$\\[6pt]
$\chi^2$/dof & $43.0/51$ & $30.7/48$\\[6pt]
\tableline\\[-3pt]
$L_{\rm x, 0.3-10}$\tablenotemark{e} &$2.0^{+0.3}_{-0.3} \times 10^{39}$  
	& $2.3^{+0.3}_{-0.3} \times 10^{39}$ \\[6pt]
$L_{\rm x, 0.3-1.5}$\tablenotemark{f} & $1.3^{+0.1}_{-0.1} \times 10^{39}$ 
	& $1.6^{+0.1}_{-0.1} \times 10^{39}$\\[6pt]
$L_{\rm x, th}$\tablenotemark{g} &$1.0^{+0.1}_{-0.1} \times 10^{39}$  
	& $1.3^{+0.1}_{-0.1} \times 10^{39}$\\[6pt]
\tableline
\end{tabular}
\tablenotetext{a}{From Dickey \& Lockman (1990).}
\tablenotetext{b}{Fixed to the typical LMXB value of 1.7.}
\tablenotetext{c}{Units of ph. keV$^{-1}$ cm$^{-2}$ s$^{-1}$.}
\tablenotetext{d}{Apec/vapec model normalization $K_{\rm vapec} =
10^{-14}/\{4\pi\,[d_A\,(1+z)]^2\}\, 
\int n_e n_H {\rm{d}}V$, where $d_A$ is 
          the angular size distance to the source (cm), $n_e$ is the electron 
          density (cm$^{-3}$), and $n_H$ is the hydrogen density (cm$^{-3}$).}
\tablenotetext{e}{Total unabsorbed luminosity of the unresolved emission 
in the $0.3$--$10$ keV band; units of erg s$^{-1}$.}
\tablenotetext{f}{Total unabsorbed luminosity of the unresolved emission 
in the $0.3$--$1.5$ keV band; units of erg s$^{-1}$.}
\tablenotetext{g}{Unabsorbed luminosity of the diffuse thermal-plasma  
component in the $0.3$--$1.5$ keV band; units of erg s$^{-1}$.}
\end{center}
\end{table}

\begin{table}
\begin{center}
\caption{Best-fit parameters for the spectrum of the unresolved 
X-ray emission in NGC\,3377. Spectral model: {\tt wabs*(apec+po)}.
Values in brackets were kept fixed. Errors are 90\% 
confidence levels for one interesting parameter ($\Delta \chi^2 =
2.7$). Parameters and units defined as in Table 4.
\label{tbl-6}}
\begin{tabular}{l@{\ \ \ \ \ }r}
\tableline\tableline\\[-6pt]
Parameter & Value\\
\tableline\\[-3pt]
$n_{\rm H}$\tablenotemark{a} & $(2.9)$ \\[6pt]
$\Gamma$  & $(1.7)$ \\[6pt]
$N_{\rm {pl}}$ at 1 keV 
	& $3.1^{+2.7}_{-2.5}\times 10^{-6}$ \\[6pt]
$kT_{\rm{gas}}$ (keV) & $0.57^{+0.19}_{-0.27}$ \\[6pt]
$Z (Z_{\odot})$ & $(1.0)$ \\[6pt]
$K_{\rm{apec}}$ &  $2.5^{+1.7}_{-1.5} \times 10^{-5}$\\[6pt]
$\chi^2$/dof & $1.4/5$ \\[6pt]
\tableline\\[-3pt]
$L_{\rm x, 0.3-10}$ 
	& $4.2^{+3.2}_{-2.2} \times 10^{38}$ \\[6pt]
$L_{\rm x, 0.3-1.5}$ 
	& $2.0^{+1.0}_{-0.8} \times 10^{38}$ \\[6pt]
$L_{\rm x, th}$ 
	& $1.0^{+1.4}_{-0.5} \times 10^{38}$ \\[6pt]
\tableline
\end{tabular}
\tablenotetext{a}{fixed to the Galactic line-of-sight value, units of
$10^{20}$ cm$^{-2}$; from Dickey \& Lockman (1990).}
\end{center}
\end{table}

%%%\clearpage
%\begin{table*}
%\begin{center}
%\caption{Fitting parameters of the isothermal $\beta$-model 
%for the extended X-ray emission, and central gas 
%densities. Errors are 90\% 
%confidence levels for one interesting parameter ($\Delta \chi^2 = 2.7$)
%\label{tbl-4}}
%\begin{tabular}{lcccccccr}
\begin{deluxetable*}{lccc@{\hspace{-0.2cm} }ccccr}
\tabletypesize{\scriptsize}
%\rotate
\tablecaption{Fitting parameters of the isothermal $\beta$-model 
for the extended soft X-ray emission, and central gas 
densities. Errors are 90\% 
confidence levels for one interesting parameter ($\Delta \chi^2 = 2.7$)
\label{tbl-4}}
\tablewidth{0pt}
\tablehead{
\colhead{} &&&&&&&&\\[6pt]
\colhead{Galaxy} & \colhead{Det. limit} & \colhead{Counts} 
	& \colhead{$S'_0$} & \colhead{$S_0$} 
	& \multicolumn{2}{c}{$r_c$} & \colhead{$\beta$} & 
	\colhead{$n_0$}\\
 & \colhead{(erg s$^{-1}$)} & \colhead{($0.3$--$1.5$ keV)} & 
	\colhead{(ct s$^{-1}$ arcsec$^{-2}$)} 
	& \colhead{(erg cm$^{-2}$ s$^{-1}$ arcsec$^{-2}$)} 
	& \colhead{(arcsec)} & \colhead{(pc)} & 
	& \colhead{(cm$^{-3}$)}}
\startdata
\, & \, & & & & & & & \\[6pt]
N821\tablenotemark{a} & $2 \times 10^{38}$
	& \multicolumn{6}{c}{Too faint for spatial analysis} 
	& $0.03^{+0.07}_{-0.02}$\\[6pt]
N3377 & $4 \times 10^{37}$ & $152 \pm 18$  
	& \hspace{-0.2cm}$(15.9 \pm 3.1) 10^{-6}$ & $(3.8\pm0.8) 10^{-17}$ &
	$4.0\pm0.5$ & $215\pm25$ & $0.50^{+0.03}_{-0.02}$
	& $0.022\pm0.004$\\[4pt]
N4486B & $1 \times 10^{38}$ 
& \multicolumn{6}{c}{No significant detection of unresolved emission} 
	& $ < 0.01$\\[4pt]
N4564 &$2 \times 10^{38}$ & $25\pm8$& \hspace{-0.2cm}$(7.7\pm4.2) 10^{-6}$ & 
	$(2.1\pm1.3) 10^{-17}$ & $8.7^{+2.3}_{-2.8}$ & $630^{+170}_{-200}$ 
	&$0.76^{+0.36}_{-0.14}$ & $0.011\pm0.006$ \\[4pt]
N4697\tablenotemark{b} 
	& $(1 \times 10^{37})$\tablenotemark{c} 
	& $1038\pm62$ & \hspace{-0.2cm}$(19.2\pm1.7) 10^{-6}$& 
	$(9.3\pm1.2) 10^{-17}$ & $6.1\pm1.7$ & $344\pm100$ 
	& $0.40^{+0.01}_{-0.01}$
	& $0.017\pm0.003$\\[4pt]
N4697\tablenotemark{d} 
	& $(1 \times 10^{37})$\tablenotemark{c} 
	& $2167\pm92$ & \hspace{-0.2cm}$(14.0\pm1.0) 10^{-6}$& 
	$(9.4\pm1.2) 10^{-17}$ & $4.6\pm1.7$ & $260\pm100$ 
	& $0.37^{+0.01}_{-0.01}$
	& $0.019\pm0.003$\\[4pt]
N5845 & $3 \times 10^{38}$
	& $55\pm12$& \hspace{-0.2cm}$(10.7\pm3.6) 10^{-6}$& 
	$(3.3\pm1.0) 10^{-17}$&$5.5\pm1.1$&$690 \pm150$&$0.59^{+0.10}_{-0.06}$
	& $0.011\pm0.004$\\
\enddata
%\tablecomments{}
%\tableline
%\end{tabular}
\tablenotetext{a}{See Fabbiano et al.~2004.}
\tablenotetext{b}{2000 Jan observation.}
\tablenotetext{c}{Detection limit for the point sources from all datasets
combined.}
\tablenotetext{d}{Combined 2003--2004 observations.}
%\tablenotetext{c}{Another sample footnote for table~\ref{tbl-2}}
%\tablecomments{We can also attach a long-ish paragraph of explanatory
%material to a table.}
%\end{center}
\end{deluxetable*}
%%%\clearpage

\subsection{Density of the hot ISM}

The gas surface brightness $S(r)$ is the projection on the sky 
of the plasma emissivity $\epsilon \approx n_e^2 \Lambda(T_{\rm {gas}})$ 
where $n_e$ is the gas electron density and $\Lambda(T_{\rm {gas}})$ 
the cooling function at a gas temperature $T_{\rm {gas}}$. 
We can obtain the gas 
density $n_e$ by de-projecting the surface brightness, 
if we know the radial dependence 
of $T_{\rm {gas}}$. The problem can be simplified 
by assuming that the temperature is constant, 
and that the radial profile  
of the surface brightness can be parameterized  
with isothermal $\beta$-models (Cavaliere \& Fusco-Femiano 
1976, 1978) of the form:
\begin{equation}
S(r) = S_0 \left[1+ \left(r/r_c\right)^2\right]^{0.5-3\beta},
\end{equation}
where the fitting parameters are $S_0$ (central surface 
brightness), $r_c$ (core radius) and $\beta$.
Physically, it can be shown that 
$\beta \sim (\sigma^2/T_{\rm {gas}})$ 
where $\sigma$ is the stellar velocity 
dispersion (e.g., Ettori 2000 and references therein).
In this approximation, 
one can derive an analytical expression for the density profile 
(Ettori 2000):
\begin{equation}
n_e(r) = n_0 \left[1+ \left(r/r_c\right)^2\right]^{-3\beta/2},
\end{equation}
where 
\begin{equation}
n_0^2 = E_0\,\pi^{-0.5}\,\frac{\Gamma(3\beta)}{\Gamma(3\beta -
0.5)}\,\frac{1}{r_c\,\Lambda(T_{\rm {gas}})}.
\end{equation}
In equation~3, $\Gamma(x)$ is the gamma function, and 
the central surface emissivity $E_0$ (luminosity 
per unit surface area of the emitter, units of 
erg s$^{-1}$ cm$^{-2}$) is simply related to the central 
surface brightness $S_0$ (received flux from a unit 
angular element on the sky, units of 
erg s$^{-1}$ cm$^{-2}$ arcsec$^{-2}$) by a constant 
geometrical factor\footnote{To obtain 
this conversion factor, recall that the area of an element of sky 
seen under a solid angle of $1$ arcsec$^2$ corresponds 
to a physical size of $(48.5 {\rm{pc}})^2 \times 
(d/10{\rm{Mpc}})^2 = 2.35 \times 10^{-11} (d/{\rm{cm}})^2$ cm$^2$, 
where $d$ is the distance to the source. The flux $S$ 
received from a unit angular element of sky 
is $1/(4\pi d^2)$ times the luminosity emitted from the 
same unit angular element, or 
$(2.35 \times 10^{-11}d^2)/(4\pi d^2)$ times the luminosity $E$ 
emitted from a unit surface area of the source.
Inverting this relation gives the numerical factor 
between surface emissivity and surface brightness 
used in the text.}, 
$E_0 = 5.34 \times 10^{11} S_0$.

In the case of hot gas in elliptical galaxies, 
the gas temperature varies at most by a factor $\approx 2$ 
over the extent of the diffuse emission 
(e.g., Sivakoff, Sarazin \& Carlin 2004; 
Sivakoff et al.~2003), while the gas 
density can vary by a few orders of magnitude.
Hence, taking into account the limited number 
of detected counts, we can accept the isothermal model 
as a good approximation for our case.
We have now all the elements 
to determine the radial density profile $n(r)$ 
and central density $n_0$ from equations 1--3, 
in the isothermal $\beta$-model approximation.
For each galaxy, we use the surface brightness 
profile inferred from the count rates of the unresolved 
emission, as described earlier, together with the cooling 
function $\Lambda(T)$ from Sutherland \& Dopita (1993).
The results are given in Table 6, 
and the inferred radial profiles of the electron density 
(equation 2) are plotted in Figure 13.

%%%\clearpage
\begin{figure}[t]
%\epsscale{1.0}
\includegraphics[angle=270,scale=0.36]{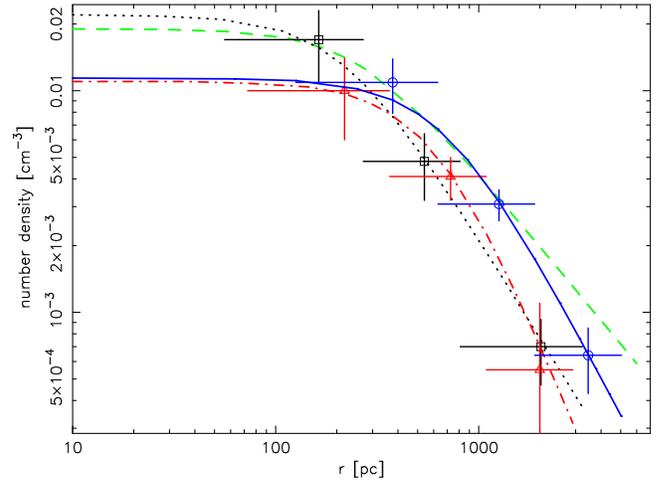}
%\includegraphics[angle=270,scale=0.36]{rprofiles_final3_dens.ps}
%\plotone{surf_bright.ps}
\caption{Density profiles of the X-ray emitting gas, 
as a function of projected physical distance from the central BH. 
Dotted (black) line: NGC\,3377; dash-dotted (red) line: 
NGC\,4564; dashed (green) line: NGC\,4697; solid (blue) 
line: NGC\,5845. The four curves have been obtained 
by deconvolving the best-fit $\beta$-model profiles.
In addition, for three of the four galaxies, 
we overplot the density directly estimated 
from the X-ray emission in three concentric shells, 
using an onion-skin approach, with no model assumptions. 
$\beta$-models and empirical onion-skin approach 
give a consistent estimate of the hot-gas density 
at distances $\ga 150$ pc; inside that radius, 
the degree of flattening in a core is somewhat 
model-dependent.\label{fig9}}
\end{figure}
%%%\clearpage
The above method for the deprojection of the surface 
brightness is based on the assumption of a specific 
analytic form for the radial profile ($\beta$-model). 
It is also possible to obtain an estimate of the 
density profile without assuming any models, 
using an onion-skin approach\footnote{An onion-skin 
approach consists in dividing 
the observed region into concentric shells, 
projected as concentric annuli on the sky.
The problem is solved from outside in. 
From the measured average temperature 
and luminosity of the gas in the outermost 
annulus, one calculates the average gas density 
in the outermost shell. The observed emission 
from the second annulus is given by emission 
from the second shell, plus some emission 
from the fraction of the outermost shell 
projected in front of it; the latter component 
is by now known, and can be subtracted. 
And so on, until the innermost annulus, 
whose emission comes from the innermost 
shell plus all other shells projected 
in front of it.} together with 
the best-fit emission measure from the fitted 
spectral model in {\footnotesize{XSPEC}} 
(recalling that $K_{\rm{apec}} \sim \int n_e^2 {\rm{d}}V$). 
%(ADD REF TO FABBIANO. ADD EXPLANATIONS?)\\
We did so for three galaxies in our 
sample\footnote{in NGC\,4697, the diffuse 
emission extends to much larger distances; hence this 
onion-skin method would give meaningful results 
only by using a larger number of shells.
A more detailed analysis of the radial profile 
of the diffuse emission in NGC\,4697 
is beyond the scope of this work, which 
focuses on the central gas density; see also 
Sarazin et al.~(2001).}: we 
defined three concentric shells, 
including the whole surface where diffuse emission 
is detected (i.e., to a radial distance of $\approx 60\arcsec$), 
and calculated the average count rate and luminosity 
from each one. From the best-fit emission measure, 
we then inferred the average gas density in the 
three shells. The results are in agreement 
with the $\beta$-model estimates (Figure 13), 
implying gas densities $\sim 1$--$2 \times 10^{-2}$ 
cm$^{-3}$ at $\sim 100$ pc from the SMBH. 

Estimating the hot-gas density in the two remaining 
galaxies (NGC\,821 and NGC\,4486B) is more difficult, 
because of the limited number of photons.
For NGC\,821, an average density 
$n_e = 4.1_{-1.5}^{+10.9} \times 10^{-3}$ cm$^{-3}$ 
was determined in Fabbiano et al.~(2004).
However, this value was estimated 
over a region of radius $20\arcsec$, excluding the point 
sources. Therefore, it underestimates the central density.
By scaling this value in analogy with the density profiles 
inferred for the other four galaxies, 
we estimate that it corresponds to central 
densities $n_e = 0.03_{-0.02}^{+0.07}$ cm$^{-3}$ 
(i.e., $\log n_e = -1.5 \pm 0.5$).
For NGC\,4486B, a central density $\approx 10^{-2}$ 
cm$^{-3}$ is the upper limit.

For the typical values of ISM temperature and density 
in our target galaxies, 
the characteristic radiative cooling timescale inside 
$r_{\rm acc}$ is $\sim 10^8$ yr, 
in the absence of external heating. 
In fact, both Type-Ia SNe and feedback 
from the SMBH are possible heating sources, 
as we shall discuss in Section 6.2.
The radiative cooling timescale is much longer 
that the accretion timescale ($\sim$ 
free-fall timescale), which is $\la 10^5$ yr 
within the sphere of influence of the SMBH 
(defined as in Section 4.3).

%\subsection{Black Hole masses}

%NGC\,821 5.0 \times 10^7 stellar kin  
%	(5.1 \pm 2.0) 10^7	(13.2+10.2-5.8) 10^7
%NGC\,3377 1.0 \times 10^8 stellar kin 
%	(10.3 +16.0-4.1) 10^7	(8.9+6.9-4.9) 10^7
%Gebhardt, K., et al. 2000, AJ, 119, 1157

%NGC\,4486B 6.0+3-2 \times 10^8 stellar kin
%		(2.0+1.5-0.9) 10^7
%Kormendy, J., et al. 1996, ApJ, 473, L91
%Kormendy, J., et al. 1997, ApJ, 482, L139

%NGC\,4564 5.7 \times 10^7 stellar kin
%	(5.7+1.3-1.7) 10^7	(1.9+1.5-0.8) 10^7
%NGC\,4697 1.2 \times 10^8 stellar kin
% ??
%NGC\,5845 3.2 \times 10^8 stellar kin
%	(35.2+20.0-7.2) 10^7	(7.1+5.5-3.1) 10^7
%Gebhardt, K., et al. 2000, AJ, 119, 1157

\subsection{Bondi inflow rate of hot gas}

From the hot-gas density (determined in Section 4.2), 
we can estimate the accretion rate 
onto a point-like mass within a uniform gas, 
in the standard model 
of steady, spherically symmetric accretion (Bondi 1952; 
Frank, King \& Raine 2002; Ruffert \& Arnett 1994):
\begin{equation}
\dot{M}_{\rm B} = \pi G^2 M^2_{\rm BH}
\frac{\rho(\infty)}{c^3_{\rm s}(\infty)} 
\left(\frac{2}{5-3\gamma}\right)^{(5-3\gamma)/(2\gamma -2)}, 
\end{equation}
where $\gamma$ is the polytropic index ($1 \le \gamma \le 5/3$), 
$c_{\rm s} = (\gamma kT/\mu m_{\rm p})^{1/2}$ is the sound
speed of the gas, $m_{\rm p}$ is the proton mass, 
$\mu m_{\rm p} $ is the mean mass per particle of gas.
We take $\mu = 0.62$ (solar composition) 
and $\gamma = 1.33$ for consistency with Pellegrini (2005).
For the gas temperature and density ``at infinity'' 
we take their values at the Bondi accretion radius 
$r_{\rm acc} = 2GM_{\rm BH}/c^2_{\rm s}(\infty)$, 
which defines the region inside which the gravitational 
pull of the SMBH dominates over the internal thermal 
energy of the gas. Henceforth, we shall refer 
to this region simply as the SMBH ``sphere of influence''
\footnote{It is also possible to define a dynamical sphere 
of influence, with a radius $\sim GM_{\rm BH}/\sigma^2$, 
where $\sigma$ is the stellar velocity dispersion. In practice, 
the two definitions are very similar for this class of systems, 
because the stellar velocity dispersion $\approx$ 
thermal velocity of virialized hot gas.}.

A spherically-symmetric model is probably 
not applicable to the behaviour of the 
accretion flow very close to the SMBH 
(e.g., Frank et al.\ 2002; Krumholz, McKee \& Klein~2005, 2006); 
however, we can still take the Bondi rate from equation~(4) as 
a first-order approximation of the rate 
at which the hot ISM is captured by the sphere of influence 
of the BH, at $r =r_{\rm acc}$. As we shall  
discuss later, not all of the inflowing, hot gas 
reaches the BH, and, conversely, there are  
other sources of gas inside $r_{\rm acc}$.

We can more usefully express the accretion radius 
and Bondi rate as: 
\begin{equation}
r_{\rm acc}  = 8.4\,\left(\frac{M_{\rm BH}}{10^8 M_{\odot}}\right)\,
\left(\frac{0.5 {\rm ~keV}}{kT}\right)\ {\rm pc}, 
\end{equation}
\begin{eqnarray}
\dot{M}_{\rm B} &=& 1.56 \times 10^{-5} 
	\left(\frac{M_{\rm BH}}{10^8 M_{\odot}}\right)^2\, 
	\left(\frac{0.5 {\rm ~keV}}{kT}\right)^{3/2}\nonumber\\
	&&\times \left(\frac{n_0}{0.01 {\rm ~cm}^{-3}}\right) \ \
	M_{\odot}\ {\rm yr}^{-1}. 
\end{eqnarray}
The numerical values for our target galaxies 
are listed in Table 7, Cols.~3,4,5
For the hot gas phase, we used a temperature 
of $0.46$ keV for NGC\,821 (Fabbiano et
al.\ 2004), $0.57$ keV for NGC\,3377, 
$0.33$ keV for NGC\,4697, and $0.5$ keV for the other two galaxies 
with a detected hot-gas component.

\subsection{Is the central gas density underestimated?}

The hot-gas densities inferred from the deprojection 
of the diffuse X-ray emission in our target galaxies 
are comparable (though, on the lower end) with 
those found in Pellegrini's (2005) sample 
($n_e \sim 0.01$--$0.5$ cm$^{-3}$)
and with the hot-gas density recently 
estimated for NGC\,3379 ($n_e \sim 0.01$ cm$^{-3}$: 
David et al.\ 2005).
There are only two nearby galaxies 
for which the accretion radius has been 
clearly resolved from {\it Chandra} images.
%: the Milky Way ($r_{\rm acc} = 1\farcs8$), 
%M\,31 ($r_{\rm acc} = 0\farcs9$), 
%and M\,87 ($r_{\rm acc} = 1\farcs9$).
For the Milky Way ($r_{\rm acc} = 1\farcs8$), 
the hot gas density $n_e(r_{\rm acc}) \sim 10^2$ cm$^{-3}$
(Baganoff et al.~2003; Quataert 2003; Cuadra et al.~2005). 
For M\,87 ($r_{\rm acc} \approx 1\farcs9$), 
the central density is $\approx 0.17$ 
cm$^{-3}$ (Di Matteo et al.~2003).
Four more galaxies have accretion radii 
$r_{\rm acc} \approx 1\arcsec$: 
in M\,31, a central density $n_e(r<200~{\rm pc}) \sim 0.1$
cm$^{-3}$ has been estimated (Garcia et al.~2005; Dosaj 
et al.~2002); $n_e \approx 0.04$ cm$^{-3}$ in the center 
of NGC\,5128 (Evans et al.~2004; Kraft et al.~2003); 
$n_e \sim 0.1$ cm$^{-3}$ for the Sombrero galaxy 
(Pellegrini et al.~2003a); and  
$n_e \sim 0.5$ cm$^{-3}$ for NGC\,4649 
(Soldatenkov, Vikhlinin \& Pavlinsky ~2003).

Our target galaxies are slightly more distant 
and/or less massive, and, as a result, their accretion radii 
are not resolved. The observed radial profiles 
provide a good estimate of the gas distribution 
at distances $\ga 150$ pc; 
however, we do not have any direct measurements 
at smaller radii (Figures 9, 13), because 
of a lack of spatial resolution. 
%(we also had 
%to remove the nuclear source from our study of the 
%diffuse emission. 
If we want to determine the amount 
of gas available for accretion inside the sphere of influence 
of the SMBH, we need to estimate the density 
at radii $\sim 10$ pc.
If $\beta$ models are a good approximation, 
the presence of a core implies that the density 
stays approximately constant inside the inner 
$r \sim 100$ pc, making this task easier. 
But we cannot {\it a priori} rule out  
the possibility that the gas density distribution 
has a cusp in the nuclear region, 
for example in correspondence of a massive star 
cluster, or because cooling gas has accumulated 
there. Although we cannot resolve 
the nuclear gas emission in the {\it Chandra} images, 
we know that it does not dominate the point-like 
nuclear sources, which have harder, AGN-like spectra (see, e.g., 
the X-ray colors in Figure 3).
A conservative constraint is that the thermal emission 
inside $\sim 10$ pc has an X-ray luminosity 
$\la 10^{38}$ erg s$^{-1}$. For typical emissivities 
corresponding to temperatures $\sim 0.3$--$0.6$ keV, 
this gives an order-of-magnitude constraint on the 
``true'' central density $n_e \la 10$ cm$^{-3}$ for 
the five target galaxies in which hot gas is found.

Moreover, the hot, X-ray emitting ISM may represent  
only a small fraction of the gas in the inner region.
That can be the case in systems where gas can cool efficiently 
(cooling timescale $<$ accretion timescale), 
or, vice versa, if gas is injected into the inner regions 
in a cool or warm phase (for example, 
through stellar winds) and is accreted 
before it has time to virialize.
Cool and warm gas would of course elude an X-ray 
investigation.
Therefore, the Bondi inflow rate of the hot-ISM component 
into the sphere of influence of the SMBH represents 
a firm lower limit to the total gas mass available for accretion.
We shall compare the SMBH luminosity expected 
for gas accretion at the Bondi rate, in the case 
of a radiatively-efficient inflow (standard disk, 
Shakura \& Sunyaev 1973) 
and for radiatively inefficient solutions.  
Conversely, we shall also estimate the accretion rate 
(expressed as a fraction of the Bondi rate) 
required to explain the observed luminosities, if 
we assume a radiatively-inefficient scenario.
In Paper II, we shall provide a complementary 
estimate of the gas injection rate into 
the SMBH sphere of influence, based 
on optical studies of the stellar population 
inside the accretion radius, following the example 
of Fabbiano et al.~(2004).

\section{Discussion}

\subsection{Accretion rates and efficiencies}

Let us define a total accretion power
$P_{\rm acc} \equiv \eta' \dot{M} c^2$, 
where $\eta'$ is the total accretion 
efficiency (radiative plus mechanical), and 
$\dot{M}$ is the rate at which matter actually 
accretes onto the BH.
One immediate objective of our analysis 
is to determine whether $\dot{M}_{\rm B}$ 
derived from X-ray observations of hot gas 
is a good tracer of $\dot{M}$.
%$\dot{M}$ receives contributions 
%from both the hot gas captured at the Bondi radius 
%(Eq.~4), and the additional 
%gas injection from stellar-mass losses (Eq.~XX).
%In general, $\dot{M} \le (\dot{M}_{\rm B} + \dot{M}_{\ast})$
%because only a fraction of gas available 
%may reach the BH; different accretion models 
%predict different values for this fraction.
The efficiency $\eta' \la 0.3$, with its exact 
value depending on the physics of the accretion flow 
as well as on the BH spin parameter; 
more generally, we shall talk of ``efficient'' 
accretion when $\eta' \sim 0.1$.
We then define a bolometric luminosity  
$L_{\rm bol} = \eta \dot{M} c^2$, where 
$\eta$ is the radiative efficiency. In general, 
it is  $\eta \equiv f_{\rm r}\eta' \la \eta'$
where $f_{\rm r}$ is the fraction 
of the accretion power released as radiation.
In standard accretion models, e.g., for 
an accretion disk extended down 
to the innermost stable orbit around the BH, 
$\eta \approx \eta' \sim 0.1$; however, 
it is possible that $\eta \ll \eta'$ 
if most of the accretion power is released, 
for example, in a jet. 
Finally, only a fraction $f_{\rm X} < 1$ 
of the bolometric luminosity 
is emitted in the {\it Chandra} X-ray band, 
$0.3$--$10$ keV; for accreting SMBHs in typical AGN, 
$f_{\rm X} \sim 10\%$ (Elvis et al.~1994; Ho 1999).
We shall also define a dimensionless accretion rate 
$\dot{m}\equiv 0.1 \dot{M} c^2/L_{\rm{Edd}}$, 
so that 
\begin{equation}
L_{\rm bol}/L_{\rm{Edd}} \equiv (\eta/0.1) \dot{m}.
\end{equation}
Similarly, we define
$\dot{m}_{\rm B} \equiv 0.1 \dot{M}_{\rm B} c^2/L_{\rm{Edd}}$. 
%and $\dot{m}_{\rm t} \equiv 
%0.1 (\dot{M}_{\rm B}+ \dot{M}_{\ast})c^2/L_{\rm{Edd}}$.

From the {\it Chandra} observations, the X-ray 
luminosity of the nuclear sources in our target galaxies 
is extremely sub-Eddington:
$L_{\rm X}/L_{\rm{Edd}}  = f_{\rm X} L_{\rm bol}/L_{\rm{Edd}}  
\sim 10^{-8}$--$10^{-7}$ (Table 7).
From the estimated densities of the hot ISM, we infer 
dimensionless Bondi accretion rates 
$\dot{m}_{\rm B}  \sim 10^{-5}$ 
(Table 7). 
%This is even more evident since we have shown that 
%the hot ISM is only a lower 
%limit to the total gas available;   
%the dimensionless total gas injection rate 
%is $\dot{m}_{\rm t}  \sim 10^{-4}$--$10^{-3}$ 
%(Tables 4, 7). 

Followig the example of previous studies (e.g., Pellegrini 2005), 
we can compare the observed X-ray luminosities 
with the inferred Bondi inflow rates. 
For the nuclei of our target galaxies, 
the X-ray observations imply the following constraint: 
\begin{equation}
\frac{L_{\rm{X}}}{0.1\dot{M}_{\rm{B}}c^2} 
	= f_{\rm X} f_{\rm r} \left(\frac{\eta'}{0.1}\right) 
	\left(\frac{\dot{m}}{\dot{m}_{\rm B}}\right) 
	\sim 5 \times 10^{-4}{\rm \,-\,}10^{-2}
\end{equation} 
(Table 7, Col.~8).
%We suggest that the physical interpretation 
%of the faint emission becomes clearer 
%if we compare the X-ray luminosity 
%with the total gas injection instead, wherever possible:
%\begin{equation}
%\frac{L_{\rm{X}}}{0.1\left(\dot{M}_{\rm{B}}+\dot{M}_{\ast}\right)c^2} 
%	= f_{\rm X} f_{\rm r} \left(\frac{\eta'}{0.1}\right) 
%	\left(\frac{\dot{m}}{\dot{m}_{\rm t}}\right) 
%	\sim 10^{-5}{\rm \,-\,}10^{-4}.
%\end{equation}
Theoretical arguments and observational evidence in a variety 
of accreting galactic nuclei suggests that $f_{\rm X}$ 
varies at most within a factor of $\approx 2$--$3$.
The crucial problem is to constrain the 
other three parameters  
($f_{\rm r}$, $\eta'$ and $\dot{m}/\dot{m}_{\rm B}$) 
in equation~(8), to make them consistent with the observed 
SMBH luminosities. This will be a test for the predictions 
of different accretion scenarios.
We shall take into account both our new sample 
of six galaxies, and another eighteen galaxies 
for which Bondi rates and SMBH luminosities 
have been measured or constrained 
(Pellegrini 2005; Garcia et al.~2005; 
David et al.~2005).  

%%%\clearpage
%\begin{table*}
%\footnotesize{
%\begin{center}
%\caption{Physical properties of the nuclear BHs ($L_{\rm X}$ refers to 
%$0.3$--$10$ keV band)\label{tbl-7}}
%\begin{tabular}{lcccccccr}
\begin{deluxetable*}{lccccccccc}
%\tabletypesize{\scriptsize}
\tablewidth{0pt}
%\rotate
\tablecaption{Physical properties of the nuclear BHs ($L_{\rm X}$ is  
the unaborbed luminosity in the $0.3$--$10$ keV band)\label{tbl-7}}
%\tablewidth{0pt}
%\tablehead{
%\hline\hline\\[-6pt]
\tablehead
{
\colhead{\,}&\colhead{\ }&\colhead{}&\colhead{}
	&\colhead{}&\colhead{}&&&&\\[8pt]
\colhead{Galaxy} & \colhead{$M_{\rm{BH}}$} & \colhead{$r_{\rm acc}$} 
        & \colhead{$\dot{M}_{\rm{B}}$} & 
	\colhead{$\log 
	\left(\frac{\dot{M}_{\rm{B}}}{\dot{M}_{\rm{Edd}}}\right)$}
	& \colhead{$\log L_{\rm{X}}$} 
        & \colhead{$\log 
	\left(\frac{L_{\rm{X}}}{L_{\rm{Edd}}}\right)$} 
        & \colhead{$L_{\rm{X}}/(0.1\dot{M}_{\rm{B}}c^2)$} 
	& \colhead{$\log L_{\rm X,ADAF}$}
	& \colhead{$\dot{M}/\dot{M}_{\rm B}$}	\\[2pt]
  &  \colhead{($10^8 M_{\odot}$)} & \colhead{(arcsec/pc)} 
	& \colhead{($M_{\odot}$ yr$^{-1}$)} &
  & \colhead{(erg s$^{-1}$)} & & 
\colhead{$= f_{\rm X} 
	\left(\frac{\eta}{0.1}\right) 
	\left(\frac{\dot{m}}{\dot{m}_{\rm B}}\right)$} 
	& \colhead{($\dot{M} = 0.1\dot{M}_{\rm B}$)}
	&\colhead{}\\[4pt]
\colhead{(1)}  &  \colhead{(2)}  &  \colhead{(3)}  & \colhead{(4)}  &  
                  \colhead{(5)}  &  \colhead{(6)}  &  \colhead{(7)} 
		 &  \colhead{(8)} &  \colhead{(9)} &  \colhead{(10)}}
\startdata
\, & \, & & & & & & & &\\[6pt]
N821 & $0.85^{+0.35}_{-0.35}$ &$0\farcs07/7.8$ & $3.8 \times 10^{-5}$& 
$-4.70^{+1.05}_{-0.43}$  & $<38.7$  
        & $< -7.3$  &   $<6 \times 10^{-3}$  & $36.4$ & $< 2$\\[4pt]
N3377 & $1.0^{+0.9}_{-0.1}$ & $0\farcs14/7.4$ & $2.8\times10^{-5}$ 
 & $-4.92^{+0.50}_{-0.24}$ & $38.5$  
        & $-7.57^{+0.10}_{-0.39}$  &  $2.3\times10^{-3}$ & $36.0$ & 
	$1.3$  \\[4pt]
N4486B
 & $\left[6.0^{+3.0}_{-2.0}\right.$ & $0\farcs61/50.4$ &
 $<1\times10^{-3}$ & $< -4.1$ & $38.4$  
        &  $-8.5^{+0.2}_{-0.3}$ &  $> 4\times10^{-5}$  & $<38.2$ 
	&$\left.>0.1\right]$\\[3pt]
	 & $0.5^{+0.5}_{-0.2}$ & $0\farcs05/4.2$ &
$<1\times10^{-5}$ & $< -5.0$ & $38.4$  
        &  $-7.44^{+0.28}_{-0.39}$ &  $> 4\times10^{-3}$  & $<35.2$ &
	$> 2$ \\[4pt]
N4564 & $0.56^{+0.03}_{-0.08}$ & $0\farcs06/4.7$ &
$5.4\times10^{-6}$
  &$-5.38^{+0.40}_{-0.40}$ & $38.9$  
        & $-6.96^{+0.23}_{-0.22}$  &  $0.03$ & $34.3$ & $6$\\[4pt]
N4697 & $1.7^{+0.2}_{-0.1}$ & $0\farcs38/21.6$ &
$1.5\times10^{-4}$ 
 & $-4.41^{+0.11}_{-0.10}$ & $38.6$  
        & $-7.74^{+0.03}_{-0.06}$  &  $4.6\times10^{-4}$ & $37.5$ 
	& $0.3$ \\[4pt]
N5845 & $2.4^{+0.4}_{-1.4}$ & $0\farcs16/20.2$ &
	$9.9\times10^{-5}$ 
	& $-4.74^{+0.37}_{-0.44}$& $39.4$  
        & $-7.00^{+0.26}_{-0.08}$  &  $5.6\times10^{-3}$ 
	& $36.8$ & $1.4$\\[4pt]
%\hline
%\end{tabular}
%\end{center}
\enddata
\tablecomments{
Col.(1): galaxy ID. NGC\,4486B is listed twice: 
the first line (in square brackets) assumes 
the mass of Kormendy et al.~(1997); the second line 
assumes the lower mass indirectly derived in our Paper II.
Col.(2): SMBH mass, see Table 1.
Col.(3): accretion radius, from equation~5.
Col.(4),(5): Bondi inflow rates estimated from the density 
and temperature of the diffuse hot ISM. 
Col.(6),(7): unabsorbed X-ray luminosity of the nuclear source, 
in the $0.3$--$10$ keV band. 
Col.(8): comparison between observationally-determined 
X-ray luminosity and bolometric luminosity expected from standard,
radiatively-efficient accretion.
Col.(9): 
prediction for the $0.3$--$10$ keV luminosity for  
an ADAF accretion flow, with a viscosity parameter $\alpha = 0.1$;
values were suitably rescaled from the fit of Merloni et al.~(2003), 
using an accretion rate $\dot{m} \approx \alpha \dot{m}_{\rm B}$, 
as expected for an ADAF scenario.
Col.(10): accretion rates required to satisfy the observed 
(unabsorbed) X-ray luminosities, as a fraction 
of the Bondi inflow rate of hot gas.
}
\end{deluxetable*}

%\section{Radiative efficiency of the accretion}

\subsection{Advective solutions (ADAF)}

Standard disk accretion predicts
$f_{\rm r} \sim 1$, $\eta' \sim 0.1$ 
and $\dot{m}/\dot{m}_{\rm B} \sim 1$.
This is ruled out by the observations: if the gas available 
were efficiently accreted at the Bondi rate, we would expect 
X-ray luminosities $\sim 10$--$100$ times higher 
than observed in our target galaxies.
Moreover, there is a theoretical reason 
why we expect standard accretion to be ruled out:
%The first possibility is that most SMBHs are faint 
%because the radiative efficiency $f_{\rm r} \eta' \ll 0.1$.
%(and hence, also the radiative efficiency $\eta \ll 0.1$). 
%This is expected to occur 
in the limit of low accretion rates 
($\dot{m} \la 0.01$), the accretion flow cannot cool 
efficiently within the infall timescale; therefore, 
most of the gravitational energy is carried by the ions 
(protons and nuclei) 
and advected into the BH, rather than being 
transferred to the electrons and radiated away. 
This regime is known as Advection-Dominated Accretion Flow (ADAF, 
Narayan \& Yi 1994; Ichimaru 1977). 

In the simplest, self-similar ADAF model, 
the luminosity  
$L_{\rm ADAF} \approx 0.1\dot{M}c^2(\dot{m}/\alpha^2) 
\approx (\dot{m}/\alpha)^2 L_{\rm Edd}$ 
(Narayan \& Yi 1994), 
where $\alpha$ is the dimensionless viscosity 
parameter, with $0.1 \la \alpha < 1$; a value $\alpha \approx 0.1$ 
is typically assumed for galactic SMBHs.
This corresponds to a radiative efficiency 
$\eta \sim 0.1 \dot{m}/\alpha^2 \sim 10 \dot{m} < 0.1$. 
A more rigorous calculation (Merloni et al.~2003), 
taking into account synchrotron and inverse-Compton 
contributions, indicates that 
the X-ray luminosity $L_{\rm X,\,ADAF} 
\propto \dot{m}^{2.3} L^{0.97}_{\rm Edd}$. In the parameter range 
of our study, the two estimates 
agree within a factor of 3.
Moreover, in the ADAF scenario, the radial inflow velocity 
inside the Bondi radius is only $\sim \alpha c_{\rm s}$. 
For a fixed hot-gas density and temperature 
at infinity, this implies that a more accurate 
estimate of the accretion rate 
is $\dot{m} \sim \alpha \dot{m}_{\rm B} \sim 0.1 
\dot{m}_{\rm B}$ (Narayan 2002).

Using the ADAF values of efficiency and accretion rate 
(from suitable rescalings of 
the Merloni et al.~2003 model fit), 
%and taking $\dot{m}_{\rm B}$ from Eq.~(6),
we estimated the $0.3$--$10$ keV luminosity 
predicted by the advective model for our target galaxies 
(Table 7, Col.~9). 
Conversely, we also determined the accretion rates 
that would be required to reproduce the observed 
luminosities, at an assumed ADAF-like efficiency 
(Table 7, Col.~10).
We find (Table 7 and Figure 14) that, at an accretion rate  
$\dot{m} \approx 0.1 \dot{m}_{\rm B}$  
(as predicted by the ADAF model),  
the SMBH luminosities are underestimated by 
a factor of 10 (for NGC\,4697) and by three to four 
orders of magnitude for the other five galaxies. 
Conversely, we find that accretion rates 
$\dot{m} \approx 0.6$--$6 \dot{m}_{\rm B}$ 
are required to reproduce the observed luminosities, 
at an ADAF-like efficiency.
In particular, an accretion rate $\approx 0.3 \dot{m}_{\rm B}$ 
suffices to explain the X-ray luminosity of NGC\,4697, 
but rates $\dot{m} \ga \dot{m}_{\rm B}$ are required 
for the other five galaxies.
If the hot-ISM inflow was the only 
source of gas, such a high accretion rate 
would be difficult to reconcile with the ADAF prediction.
%A simplistic suggestion would be 
%that the inflow is only partly advective, 
%or has an advective and a standard component, 
%so that the average efficiency is intermediate 
%between the standard and the ADAF values. However, 
%this is inconsistent with all theoretical 
%expectations of low radiative efficiency  
%at low accretion rates.
We take this finding as a strong   
indication that, for our target galaxies, 
the gas available for accretion has been underestimated:
there are other sources of fuel  
for the SMBH, such as stellar winds, 
that could be more significant 
than the Bondi inflow; 
this would explain why we infer $\dot{m} \ga \dot{m}_{\rm B}$.
We shall discuss this scenario in detail 
in Paper II (see also Fabbiano et al.~2004; Pellegrini 2005).

%%%\clearpage
\begin{figure}
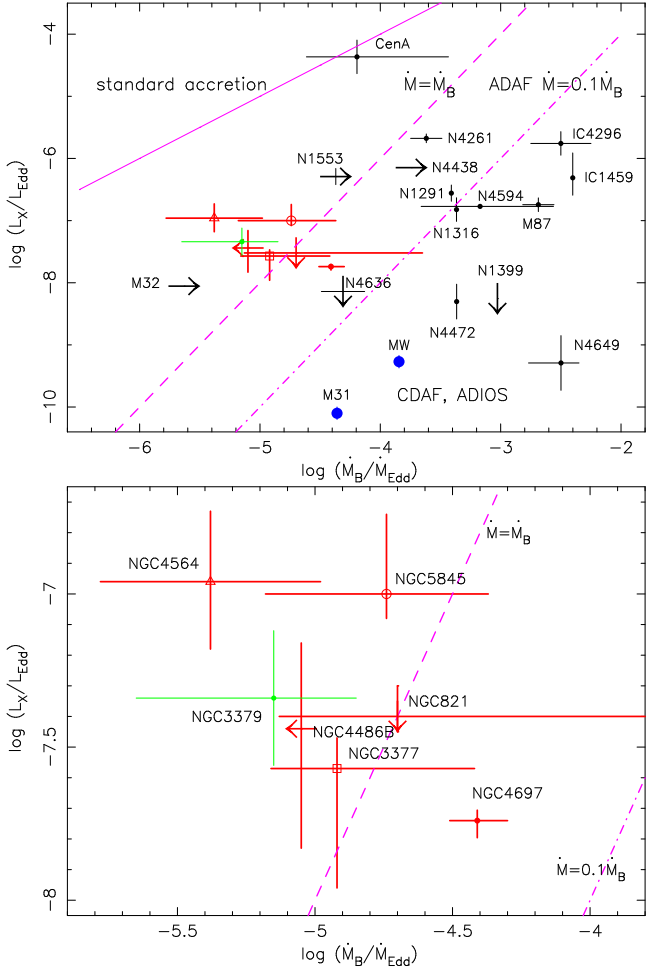

%\epsscale{1.0}
\includegraphics[angle=270,scale=0.36]{f14a.eps}\\
\includegraphics[angle=270,scale=0.36]{f14b.eps}
%\includegraphics[angle=270,scale=0.36]{silvia_1a.ps}\\
%\includegraphics[angle=270,scale=0.36]{silvia_1b.ps}
%\plotone{surf_bright.ps}
\caption{Top panel: relation between normalized X-ray luminosity 
$L_{\rm X}/L_{\rm Edd}$ and normalized Bondi accretion 
rate $\dot{m}_{\rm B} = \dot{M}_{\rm B}/\dot{M}_{\rm Edd}$.
Bottom panel: zoomed-in view of the same diagram, 
to allow a better identification of the 
six galaxies of the present study 
(datapoints plotted in red in the online version).
The datapoints 
in black have been collected by Pellegrini (2005) 
(see references therein for the original measurements; 
a slightly revised mass has been adopted for NGC\,4636, 
from Wang \& Zhang 2003).
The large filled circles (in blue, in the online version) 
represent Sag A$^{\ast}$ (Baganoff et al.~2003) 
and M\,31$^{\ast}$ (Garcia et al.~2005). The datapoint 
plotted in green is NGC\,3379 (David et al.~2005). 
We have compared the observed X-ray luminosities with 
those predicted, as a function of Bondi accretion rate,  
by: a) standard radiatively-efficient 
accretion ($\eta = 0.1$, $\dot{m} = \dot{m}_{\rm B}$: 
solid line); b) standard ADAF solution ($\eta = 10\dot{m}$, 
$\dot{m} = 0.1 \dot{m}_{\rm B}$: 
dashed-dotted line); c) a scenario 
with ADAF-like efficiency ($\eta = 10\dot{m}$) 
but higher accretion rate ($\dot{m} = \dot{m}_{\rm B}$: 
dashed line).
SMBHs with X-ray luminosities falling below the standard 
ADAF line are likely to require an additional mechanism 
to reduce the true mass accretion rate further 
below the Bondi value (e.g., ADIOS or CDAF solutions). 
Conversely, for the six target galaxies 
studied in this paper, the Bondi rate is too low  
to be consistent with the observed X-ray luminosities, 
if the accretion process has ADAF-like radiative 
efficiencies. \label{fig10}}
\end{figure}
%%%\clearpage

More generally, there are $\approx 20$ quiescent 
or very faint SMBHs for which the SMBH mass 
and X-ray luminosity, and the density of 
the surrounding hot ISM, 
have been estimated or constrained (Pellegrini 2005; 
Garcia et al.~2005; David et al.~2005). 
When plotted in an X-ray-luminosity 
versus Bondi-rate diagram (Figure 14), 
the distribution of the datapoints  
does not provide any obvious correlations 
between these two parameters, 
or between the luminosities observed and 
predicted by the ADAF model.
(The standard disk model fares much worse, 
and is clearly ruled out for all but one galaxy).
Some SMBHs, including those in the six 
galaxies of our sample, 
tend to be more X-ray luminous than predicted 
by the standard ADAF model (i.e., they lie 
above the predicted $\dot{m} \approx 0.1 \dot{m}_{\rm B}$ 
ADAF line in Figure 14). This seems to be the case 
for most galaxies with extremely low
hot-ISM densities in the nuclear region ($n_e \la 0.1$ cm$^{-3}$), 
and therefore very low Bondi accretion rates 
($\dot{m}_{\rm B} \la 10^{-4}$).
On the other hand, SMBHs in galaxies with estimated 
Bondi accretion rates 
$\dot{m}_{\rm B} \ga 10^{-4}$ tend to lie below 
the ADAF line (Figure 14); that is, the ADAF model 
overpredicts their observed luminosities, in some cases 
by more than two orders of magnitude.

In the framework of inefficient accretion, 
we can explain the former group of galaxies by considering 
additional sources of gas which boost the accretion 
rate above the classical Bondi value, as 
we shall discuss in Paper II. 
Conversely, an additional mechanism must be at work 
in the latter group of galaxies,  
reducing the true accretion rate to values 
$\dot{m} \la 10^{-2} \dot{m}_{\rm B}$.
As an aside, we note that 
the lack of SMBHs in the bottom left of the diagram 
(i.e., with $\dot{m} \la 10^{-5} 
\dot{m}_{\rm B}$ and $L_{\rm X} \la 10^{-8} L_{\rm Edd}$) 
may be partly due to a selection effect, because 
both the nuclear point-like source and the diffuse 
hot-ISM emission would be close to or 
below the {\it Chandra} detection 
limit in those galaxies.

%We shall re-discuss the relation between 
%SMBH luminosity and accretion rate in Section 7, 
%after we have suggested a simple model to take 
%into account both the accretion power and mass budget. 
%We shall propose a possible physical interpretation, based 
%on three physical parameters (X-ray luminosity, 
%accretion fraction and total gas injection rate).

\subsection{Convection and outflows}

The original ADAF solution has subsequently been developed 
by taking into account two physical effects that 
are likely to further reduce the accretion rate, so that  
$\dot{m} \ll \dot{m}_{\rm B}$ (see Narayan 2002 
and references therein, for a review). Firstly, 
the inflowing gas is likely to become convectively 
unstable due to its inefficient cooling (Convection 
Dominated Accretion Flow, CDAF: 
Narayan et al.~2000); convective instabilities 
are also likely to occur when a magnetic field 
is frozen into the infalling gas 
(Igumenshchev \& Narayan 2002); or when 
the inflowing gas has a large angular 
momentum (Proga \& Begelman 2003) 
or vorticity (Krumholz, McKee \& Klein~2005, 2006). Secondly, 
inefficient cooling causes the gas to have positive 
Bernoulli parameter, thus triggering large-scale winds 
and outflows which reduce the amount of gas reaching  
the BH (Advection Dominated Inflow-Outflow Solutions, 
ADIOS: Blandford \& Begelman 1999). 

%In our six target galaxies, very low accretion 
%rates ($\dot{m}/\dot{m}_{\rm B} < 0.1$) 
%are not needed, if the central gas densities 
%are as low as inferred from our $\beta$-model fitting: 
%a standard ADAF model suffices to account for 
%the X-ray faintness, as discussed earlier.
%However, t
%Typical values $\dot{m}/\dot{m}_{\rm B} \sim 10^{-2}$ 
%are needed for most of the galaxies with $\dot{m}_{\rm B} 
%\ga 10^{-4}$, lying below the ADAF line in Figure 10. 
%Among them, the SMBH in the Milky Way, for which 
%$\dot{m}_{\rm B} \approx 2 \times 10^{-4}$; 
%an accretion rate $\dot{m}/\dot{m}_{\rm B} \la 10^{-3}$ 
%has been suggested (Quataert 2003). 
%In terms of gas densities, we can summarize the observational results 
%by saying that the Bondi inflow rate applied to 
%radiatively-inefficient ADAF solutions 
%underpredicts the X-ray luminosities of nuclear BHs when 
%the central hot-ISM densities are $\la 0.1$ cm$^{-3}$; 
%that is, it would appear that the true accretion rate 
%is higher than estimated from the Bondi rate in those cases.
%Conversely, it may be necessary to postulate 

Low, ADAF-like efficiencies {\em and}
a reduced accretion rate with respect to the Bondi rate 
($\dot{m}/\dot{m}_{\rm B} \sim 10^{-3}$--$10^{-2}$),
via CDAF and/or ADIOS solutions, is a possible explanation 
for those galaxies in which hot-gas densities 
are $\sim 0.1$--$10$ cm$^{-3}$, 
and correspondingly, $10^{-4} \la \dot{m}_{\rm B} 
\la 10^{-2}$. They lie below the standard ADAF line 
in Figure 14. Among them, the SMBH in the Milky Way, for which 
$\dot{m}_{\rm B} \approx 2 \times 10^{-4}$; 
an accretion rate $\dot{m}/\dot{m}_{\rm B} \la 10^{-3}$ 
has been suggested (Quataert 2003).
The SMBH in the nucleus of M\,31 also 
belongs to this group (Garcia et al.~2005).
At higher central gas densities ($\ga 10$ cm$^{-3}$, 
corresponding to  
$\dot{m}_{\rm B} \la 10^{-2}$), theoretical and observational 
arguments show that accretion will become radiatively 
efficient, with the formation of a standard, optically-thick, 
geometrically-thin accretion disk. SMBHs in this regime 
will appear as fully-fledged AGN.

\subsection{Intermittent accretion}

We have assumed so far that 
the accretion rate can be expressed as a constant 
fraction of the gas available.
We also need to take into account 
the possibility that accretion is intermittent, so that 
the system may go through phases when 
the SMBH accretion rate is much lower than the Bondi 
inflow rate, and subsequent phases in which it can be 
much higher. The nuclear region would alternatively 
build up or be depleted of gas.

In one version of this scenario, it is the feedback from the nuclear 
X-ray source that drives and self-regulates the activity cycles: 
when the nuclear source is active, the surrounding ISM 
is radiatively heated (Ciotti \& Ostriker 2001; 
Ostriker \& Ciotti 2005),
or, in the case of radiatively-inefficient 
accretion, is heated by the bulk kinetic 
energy of jets, outflows or convective motions 
(Binney \& Tabor 1995; 
Omma et al.~2004), until the radial inflow is stopped 
and the nucleus switches off. Thus, the Bondi inflow 
estimated from the gas density at large radii 
(at the limit of the {\it Chandra} 
resolution) may not provide a good indication 
for the time-dependent accretion rate and SMBH 
luminosity. This is a possible explanation for 
the observed lack of any relation  
between $L_{\rm X}$, $M_{\rm BH}$ and $\dot{M}_{\rm B}$
(Pellegrini 2005).

Alternatively, accretion may cycle 
between radiatively inefficient flows (e.g., CDAF) 
in which gas accumulates inside the accretion radius, 
and a standard, radiatively-efficient accretion disk 
in which the stored gas is quickly accreted into the BH.
However, this transition should only happen 
at accretion rates $\dot{m} \sim 10^{-2}$, 
much higher than estimated in our target galaxies.

Intermittent accretion may also be due to 
thermal-viscous ionization instabilities, triggered 
by an ionization front propagating inwards or outwards 
across an accretion disk 
(Siemiginowska, Czerny \& Kostyunin 1996); a change 
of the viscosity parameter by a factor of $4$ 
between the two states may lead to changes 
in luminosity by four order of magnitudes. 
Again, this scenario requires that the inflowing gas 
can cool efficiently to form an accretion disk.

It was suggested (Fabbiano et al.~2004) 
that intermittent ejections may be responsible 
for the knotty jet-like feature in the nuclear 
region of NGC\,821. 
If this emission is connected with nuclear activity, 
the most likely explanation would be that of thermal 
emission from shocks in the hot ISM 
resulting from a past nuclear outburst.

%\section{Optical comparisons}

\section{Conclusions}

We have used {\it Chandra}, supported by archival 
{\it HST} and ground-based optical images, 
to study the nature, accretion power 
and mass budget of X-ray-faint SMBHs in elliptical/S0 galaxies. 
For this purpose, we have chosen a sample of six nearby galaxies 
with quiescent SMBHs, five of which with 
accurate BH mass determinations (from stellar kinematics).
From the X-ray images, we found different morphologies 
for the X-ray emission in the nuclear region.  
In NGC\,821, a jet-like feature has been previously 
suggested (Fabbiano et al.~2004). In NGC\,4697, 
the nuclear emission is resolved into a few point-like sources 
(one of which is coincident with the nucleus). In other cases, 
(NGC\,3377, NGC\,4564, NGC\,5845) the nuclear emission 
is extended on scales of a few hundred pc 
but its morphology remains unclear. 
Obscuration of the nuclear source may play 
a role in some cases; for example, in at least one galaxy 
(NGC\,5845), the SMBH may be partly obscured by an edge-on 
disk of dust and stars, and the observed extended X-ray 
emission may be due to scattering in the surrounding plasma 
above and below the disk plane.

In five of the six galaxies, the X-ray colors 
of the nuclear emission are hard, consistent with 
an AGN spectrum. Hence, it is likely that most of 
the nuclear emission comes from the accreting SMBH, rather than from 
a central concentration of hot gas. 
In addition, softer X-ray emission 
from diffuse hot gas is detected 
around the nuclear region, out to $\sim 1$ kpc. 
We have fitted the soft X-ray brightness profiles 
with $\beta$ models to estimate central hot gas densities 
(typical values are $\sim 0.02 \pm 0.01$ cm$^{-3}$).
For two of our galaxies, the flux from the diffuse 
component is high enough to allow 
also for spectral fitting. 
NGC\,4486B is the only case where the central, 
point-like X-ray source is soft but there is 
no evidence of diffuse thermal emission around it.

The Bondi accretion rate gives an estimate 
of the rate at which diffuse hot gas flows 
into the sphere of influence of the SMBH (characteristic 
radius $r_{\rm acc} \sim 10$ pc $\sim 0\farcs1$). 
Typical Bondi 
inflow rates vary from $\sim$ a few $\times 10^{-6}$ 
to $\sim 10^{-4} M_{\odot}$ yr$^{-1}$ in our target galaxies
This is a lower limit to the gas injection rate into the sphere 
of influence, because it does not take into account 
additional gas sources directly inside $r_{\rm acc}$ 
(for example, stellar winds or gas losses 
from Type-Ia supernovae).

Based on the X-ray data, we tried to answer the following 
(inter-related) questions:
whether the observed X-ray SMBH luminosities 
are consistent with standard or radiatively-inefficient 
accretion; whether the X-ray-inferred Bondi rate 
is a good estimate of the true accretion rate 
onto the SMBH; and whether there is a correlation 
between Bondi rate and X-ray luminosity of the SMBH.
We found that standard accretion is ruled out 
for all galaxies, as it would overestimate the SMBH 
luminosities. On the other hand, our quiescent sample of SMBHs 
are also brighter than what is predicted by 
radiatively-inefficient solutions (ADAF, ADIOS, CDAF); 
we interpret this as evidence 
that the Bondi inflow rate of the hot ISM 
underestimates the total gas injection rate 
into the sphere of influence of the SMBH.
However, when we consider the other X-ray faint SMBHs 
for which the nuclear luminosity and the Bondi 
rate are known from the literature, we see that
most of them are consistent with, or fainter than 
predicted by the ADAF model (as noted in Pellegrini 2005). 
In those cases, it appears that the Bondi inflow rate of hot gas 
is an over-estimate of the true accretion rate.
Overall, there seems to be no correlation 
between the Bondi rate and the X-ray luminosity of the SMBH.
A possible explanation suggested (Pellegrini 2005)
for this lack of correlation 
is that accretion is intermittent, self-regulated 
by feedback from the BH: it may switch 
between a state in which $\dot{m} \ll \dot{m}_{\rm B}$ 
(when gas builds up in the inner few pc), 
and one in which $\dot{m} \gg \dot{m}_{\rm B}$ 
(as it is depleted).
%without noticeable 
%differences in the gas distribution at larger 
%radii ($\ga 150$ pc), in the region 
%probed by {\it Chandra}.

In Paper II we will use optical observations 
to investigate additional sources of fuel 
for the SMBH; this will provide a better 
estimate of the total accretion rate, 
and will allow us to re-analyze its correlation 
with the X-ray luminosity. We shall then 
be able to estimate the energy and mass budget, 
discuss the conditions for a steady-state, and 
distinguish between different varieties 
of radiatively-inefficient models.

\acknowledgments

We thank Andreas Zezas for his suggestions 
on the X-ray data analysis and interpretation, 
and Hermine Landt for her detailed comments on the 
accretion rate issue. RS acknowleges 
support from University College London 
via a Marie Curie Fellowship. AB acknowledges partial support 
from NASA contract NAS8-39073 and
from NASA grants GO1-2115X and GO2-3135X.

%Larry David 
%for making a preprint of his work on NGC\,3379 available 
%before submission.

\end{document}